# First M87 Event Horizon Telescope Results. I. The Shadow of the Supermassive Black Hole

The Event Horizon Telescope Collaboration
(See the end matter for the full list of authors.)

## Abstract

When surrounded by a transparent emission region, black holes are expected to reveal a dark shadow caused by gravitational light bending and photon capture at the event horizon. To image and study this phenomenon, we have assembled the Event Horizon Telescope, a global very long baseline interferometry array observing at a wavelength of 1.3 mm. This allows us to reconstruct event-horizon-scale images of the supermassive black hole candidate in the center of the giant elliptical galaxy M87. We have resolved the central compact radio source as an asymmetric bright emission ring with a diameter of $42 \pm 3$ $\mu$as, which is circular and encompasses a central depression in brightness with a flux ratio $\gtrsim$10:1. The emission ring is recovered using different calibration and imaging schemes, with its diameter and width remaining stable over four different observations carried out in different days. Overall, the observed image is consistent with expectations for the shadow of a Kerr black hole as predicted by general relativity. The asymmetry in brightness in the ring can be explained in terms of relativistic beaming of the emission from a plasma rotating close to the speed of light around a black hole. We compare our images to an extensive library of ray-traced general-relativistic magnetohydrodynamic simulations of black holes and derive a central mass of $M = (6.5 \pm 0.7) \times 10^9\, M_\odot$. Our radio-wave observations thus provide powerful evidence for the presence of supermassive black holes in centers of galaxies and as the central engines of active galactic nuclei. They also present a new tool to explore gravity in its most extreme limit and on a mass scale that was so far not accessible.

## 1. Introduction

Black holes are a fundamental prediction of the theory of general relativity (GR; Einstein 1915). A defining feature of black holes is their event horizon, a one-way causal boundary in spacetime from which not even light can escape (Schwarzschild 1916). The production of black holes is generic in GR (Penrose 1965), and more than a century after Schwarzschild, they remain at the heart of fundamental questions in unifying GR with quantum physics (Hawking 1976; Giddings 2017).

Black holes are common in astrophysics and are found over a wide range of masses. Evidence for stellar-mass black holes comes from X-ray (Webster & Murdin 1972; Remillard & McClintock 2006) and gravitational-wave measurements (Abbott et al. 2016). Supermassive black holes, with masses from millions to tens of billions of solar masses, are thought to exist in the centers of nearly all galaxies (Lynden-Bell 1969; Kormendy & Richstone 1995; Miyoshi et al. 1995), including in the Galactic center (Eckart & Genzel 1997; Ghez et al. 1998; Gravity Collaboration et al. 2018a) and in the nucleus of the nearby elliptical galaxy M87 (Gebhardt et al. 2011; Walsh et al. 2013).

Active galactic nuclei (AGNs) are central bright regions that can outshine the entire stellar population of their host galaxy. Some of these objects, quasars, are the most luminous steady sources in the universe (Schmidt 1963; Sanders et al. 1989) and are thought to be powered by supermassive black holes accreting matter at very high rates through a geometrically thin, optically thick accretion disk (Shakura & Sunyaev 1973; Sun & Malkan 1989). In contrast, most AGNs in the local universe, including the Galactic center and M87, are associated with supermassive black holes fed by hot, tenuous accretion flows with much lower accretion rates (Ichimaru 1977; Narayan & Yi 1995; Blandford & Begelman 1999; Yuan & Narayan 2014).

In many AGNs, collimated relativistic plasma jets (Bridle & Perley 1984; Zensus 1997) launched by the central black hole contribute to the observed emission. These jets may be powered either by magnetic fields threading the event horizon, extracting the rotational energy from the black hole (Blandford & Znajek 1977), or from the accretion flow (Blandford & Payne 1982). The near-horizon emission from low-luminosity active galactic nuclei (LLAGNs; Ho 1999) is produced by synchrotron radiation that peaks from the radio through the far-infrared. This emission may be produced either in the accretion flow (Narayan et al. 1995), the jet (Falcke et al. 1993), or both (Yuan et al. 2002).

When viewed from infinity, a nonrotating Schwarzschild (1916) black hole has a photon capture radius $R_c = \sqrt{27}\, r_g$, where $r_g \equiv GM/c^2$ is the characteristic lengthscale of a black hole. The photon capture radius is larger than the Schwarzschild radius $R_S$ that marks the event horizon of a nonrotating black hole, $R_S \equiv 2\, r_g$. Photons approaching the black hole with an impact parameter $b < R_c$ are captured and plunge into the black hole (Hilbert 1917); photons with $b > R_c$ escape to infinity; photons with $b = R_c$ are captured on an unstable circular orbit and produce what is commonly referred to as the lensed "photon ring." In the Kerr (1963) metric, which describes black holes with spin angular momentum, $R_c$ changes with the ray's orientation relative to the angular-momentum vector, and the black hole's cross section is not necessarily circular (Bardeen 1973). This change is small ($\lesssim$4%), but potentially detectable (Takahashi 2004; Johannsen & Psaltis 2010).

The simulations of Luminet (1979) showed that for a black hole embedded in a geometrically thin, optically thick accretion disk, the photon capture radius would appear to a distant observer as a thin emission ring inside a lensed image of the accretion disk. For accreting black holes embedded in a geometrically thick, optically thin emission region, as in LLAGNs, the combination of an event horizon and light bending leads to the appearance of a dark "shadow" together with a bright emission ring that should be detectable through very long baseline interferometery (VLBI) experiments (Falcke et al. 2000a). Its shape can appear as a "crescent" because of fast rotation and relativistic beaming (Falcke et al. 2000b; Bromley et al. 2001; Noble et al. 2007; Broderick & Loeb 2009; Kamruddin & Dexter 2013; Lu et al. 2014).

The observed projected diameter of the emission ring, which contains radiation primarily from the gravitationally lensed photon ring, is proportional to $R_c$ and hence to the mass of the black hole, but also depends nontrivially on a number of factors: the observing resolution, the spin vector of the black hole and its inclination, as well as the size and structure of the emitting region. These factors are typically of order unity and can be calibrated using theoretical models.

Modern general-relativistic simulations of accretion flows and radiative transfer produce realistic images of black hole shadows and crescents for a wide range of near-horizon emission models (Broderick & Loeb 2006; Mościbrodzka et al. 2009; Dexter et al. 2012; Dibi et al. 2012; Chan et al. 2015; Mościbrodzka et al. 2016; Porth et al. 2017; Chael et al. 2018a; Ryan et al. 2018; Davelaar et al. 2019). These images can be used to test basic properties of black holes as predicted in GR (Johannsen & Psaltis 2010; Broderick et al. 2014; Psaltis et al. 2015), or in alternative theories of gravity (Grenzebach et al. 2014; Younsi et al. 2016; Mizuno et al. 2018). They can also be used to test alternatives to black holes (Bambi & Freese 2009; Vincent et al. 2016; Olivares et al. 2019).

VLBI at an observing wavelength of 1.3 mm (230 GHz) with Earth-diameter-scale baselines is required to resolve the shadows of the core of M87 (M87$^*$ hereafter) and of the Galactic center of Sagittarius A$^*$ (Sgr A$^*$, Balick & Brown 1974), the two supermassive black holes with the largest apparent angular sizes (Johannsen et al. 2012). At 1.3 mm and shorter wavelengths, Earth-diameter VLBI baselines achieve an angular resolution sufficient to resolve the shadow of both sources, while the spectra of both sources become optically thin, thus revealing the structure of the innermost emission region. Early pathfinder experiments (Padin et al. 1990; Krichbaum et al. 1998) demonstrated the feasibility of VLBI techniques at ~1.3 mm wavelengths. Over the following decade, a program to improve sensitivity of 1.3 mm-VLBI through development of broadband instrumentation led to the detection of event-horizon-scale structures in both Sgr A$^*$ and M87$^*$ (Doeleman et al. 2008, 2012). Building on these observations, the Event Horizon Telescope (EHT) collaboration was established to assemble a global VLBI array operating at a wavelength of 1.3 mm with the required angular resolution, sensitivity, and baseline coverage to image the shadows in M87$^*$ and Sgr A$^*$.

In this paper, we present and discuss the first event-horizon-scale images of the supermassive black hole candidate M87$^*$ from an EHT VLBI campaign conducted in 2017 April at a wavelength of 1.3 mm. The accompanying papers give a more extensive description of the instrument (EHT Collaboration et al. 2019a, Paper II), data reduction (EHT Collaoration et al. 2019b, hereafter Paper III), imaging of the M87 shadow (EHT Collaboration et al. 2019c, hereafter Paper IV), theoretical models (EHT Collaboration et al. 2019d, hereafter Paper V), and the black hole mass estimate (EHT Collaboration et al. 2019e, hereafter Paper VI).

## 2. The Radio Core in M87

In Curtis (1918), Heber Curtis detected a linear feature in M87, later called a "jet" by Baade & Minkowski (1954). The jet is seen as a bright radio source, Virgo A or 3C 274 (Bolton et al. 1949; Kassim et al. 1993; Owen et al. 2000), which extends out to 65 kpc with an age estimated at about 40 Myr and a kinetic power of about $10^{42}$ to $10^{45}$ erg s$^{-1}$ (de Gasperin et al. 2012; Broderick et al. 2015). It is also well studied in the optical (Biretta et al. 1999; Perlman et al. 2011), X-ray (Marshall et al. 2002), and gamma-ray bands (Abramowski et al. 2012). The upstream end of the jet is marked by a compact radio source (Cohen et al. 1969). Such compact radio sources are ubiquitous in LLAGNs (Wrobel & Heeschen 1984; Nagar et al. 2005) and are believed to be signatures of supermassive black holes.

The radio structures of the large-scale jet (Owen et al. 1989; de Gasperin et al. 2012) and of the core of M87 (Reid et al. 1989; Junor et al. 1999; Hada et al. 2016; Mertens et al. 2016; Kim et al. 2018b; Walker et al. 2018) have been resolved in great detail and at multiple wavelengths. Furthermore, the leveling-off of the "core-shift" effect (Blandford & Königl 1979), where the apparent position of the radio core shifts in the upstream jet direction with decreasing wavelength from increased transparency to synchrotron self-absorption, indicates that at a wavelength of 1.3 mm M87$^*$ is coincident with the supermassive black hole (Hada et al. 2011). The envelope of the jet limb maintains a quasi-parabolic shape over a wide range of distances from $\sim10^5\,r_g$ down to $\sim20\,r_g$ (Asada & Nakamura 2012; Hada et al. 2013; Nakamura & Asada 2013; Nakamura et al. 2018; Walker et al. 2018).

VLBI observations at 1.3 mm have revealed a diameter of the emission region of ~40 $\mu$as, which is comparable to the expected horizon-scale structure (Doeleman et al. 2012; Akiyama et al. 2015). These observations, however, were not able to image the black hole shadow due to limited baseline coverage.

Based on three recent stellar population measurements, we here adopt a distance to M87 of 16.8 ± 0.8 Mpc (Blakeslee et al. 2009; Bird et al. 2010; Cantiello et al. 2018, see Paper VI). Using this distance and the modeling of surface brightness and stellar velocity dispersion at optical wavelengths (Gebhardt & Thomas 2009; Gebhardt et al. 2011), we infer the mass of M87$^*$ to be $M = 6.2^{+1.1}_{-0.6} \times 10^9\,M_\odot$ (see Table 9 in Paper VI). On the other hand, mass measurements modeling the kinematic structure of the gas disk (Harms et al. 1994; Macchetto et al. 1997) yield $M = 3.5^{+0.9}_{-0.3} \times 10^9\,M_\odot$ (Walsh et al. 2013, Paper VI). These two mass estimates, from stellar and gas dynamics, predict a theoretical shadow diameter for a Schwarzschild black hole of $37.6^{+6.2}_{-3.5}$ $\mu$as and $21.3^{+5}_{-1.7}$ $\mu$as, respectively.

## 3. The Event Horizon Telescope

The EHT (Paper II) is a VLBI experiment that directly measures "visibilities," or Fourier components, of the radio brightness distribution on the sky. As the Earth rotates, each telescope pair in the network samples many spatial frequencies.

The array has a nominal angular resolution of $\lambda/L$, where $\lambda$ is the observing wavelength and $L$ is the maximum projected baseline length between telescopes in the array (Thompson et al. 2017). In this way, VLBI creates a virtual telescope that spans nearly the full diameter of the Earth.

To measure interferometric visibilities, the widely separated telescopes simultaneously sample and coherently record the radiation field from the source. Synchronization using the Global Positioning System typically achieves temporal alignment of these recordings within tens of nanoseconds. Each station is equipped with a hydrogen maser frequency standard. With the atmospheric conditions during our observations the coherent integration time was typically 10 s (see Figure 2 in Paper II). Use of hydrogen maser frequency standards at all EHT sites ensures coherence across the array over this timescale. After observations, recordings are staged at a central location, aligned in time, and signals from each telescope-pair are cross-correlated.

While VLBI is well established at centimeter and millimeter wavelengths (Boccardi et al. 2017; Thompson et al. 2017) and can be used to study the immediate environments of black holes (Krichbaum et al. 1993; Doeleman et al. 2001), the extension of VLBI to a wavelength of 1.3 mm has required long-term technical developments. Challenges at shorter wavelengths include increased noise in radio receiver electronics, higher atmospheric opacity, increased phase fluctuations caused by atmospheric turbulence, and decreased efficiency and size of radio telescopes in the millimeter and submillimeter observing bands. Started in 2009 (Doeleman et al. 2009a), the EHT began a program to address these challenges by increasing array sensitivity. Development and deployment of broadband VLBI systems (Whitney et al. 2013; Vertatschitsch et al. 2015) led to data recording rates that now exceed those of typical cm-VLBI arrays by more than an order of magnitude. Parallel efforts to support infrastructure upgrades at additional VLBI sites, including the Atacama Large Millimeter/submillimeter Array (ALMA; Matthews et al. 2018; Goddi et al. 2019) and the Atacama Pathfinder Experiment telescope (APEX) in Chile (Wagner et al. 2015), the Large Millimeter Telescope Alfonso Serrano (LMT) in Mexico (Ortiz-León et al. 2016), the IRAM 30 m telescope on Pico Veleta (PV) in Spain (Greve et al. 1995), the Submillimeter Telescope Observatory in Arizona (SMT; Baars et al. 1999), the James Clerk Maxwell Telescope (JCMT) and the Submillimeter Array (SMA) in Hawai'i (Doeleman et al. 2008; Primiani et al. 2016; Young et al. 2016), and the South Pole Telescope (SPT) in Antarctica (Kim et al. 2018a), extended the range of EHT baselines and coverage, and the overall collecting area of the array. These developments increased the sensitivity of the EHT by a factor of ~30 over early experiments that confirmed horizon-scale structures in M87* and Sgr A* (Doeleman et al. 2008, 2012; Akiyama et al. 2015; Johnson et al. 2015; Fish et al. 2016; Lu et al. 2018).

For the observations at a wavelength of 1.3 mm presented here, the EHT collaboration fielded a global VLBI array of eight stations over six geographical locations. Baseline lengths ranged from 160 m to 10,700 km toward M87*, resulting in an array with a theoretical diffraction-limit resolution of ~25 $\mu$as (see Figures 1 and 2, and Paper II).

## 4. Observations, Correlation, and Calibration

We observed M87* on 2017 April 5, 6, 10, and 11 with the EHT. Weather was uniformly good to excellent with nightly median zenith atmospheric opacities at 230 GHz ranging from 0.03 to 0.28 over the different locations. The observations were scheduled as a series of scans of three to seven minutes in duration, with M87* scans interleaved with those on the quasar 3C 279. The number of scans obtained on M87* per night ranged from 7 (April 10) to 25 (April 6) as a result of different observing schedules. A description of the M87* observations, their correlation, calibration, and validated final data products is presented in Paper III and briefly summarized here.

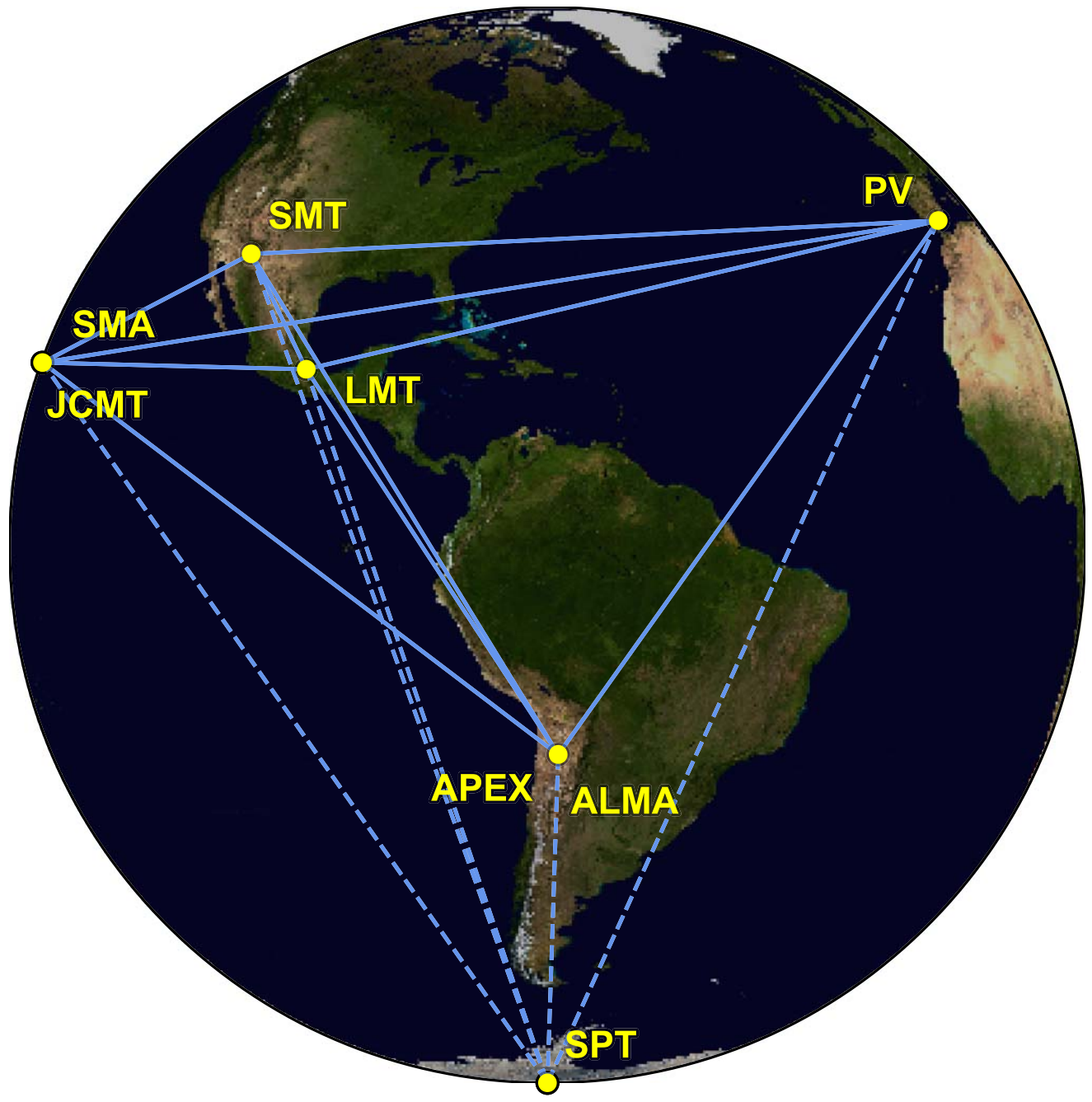

**Figure 1.** Eight stations of the EHT 2017 campaign over six geographic locations as viewed from the equatorial plane. Solid baselines represent mutual visibility on M87* (+12° declination). The dashed baselines were used for the calibration source 3C279 (see Papers III and IV).

At each station, the astronomical signal in both polarizations and two adjacent 2 GHz wide frequency bands centered at 227.1 and 229.1 GHz were converted to baseband using standard heterodyne techniques, then digitized and recorded at a total rate of 32 Gbps. Correlation of the data was carried out using a software correlator (Deller et al. 2007) at the MIT Haystack Observatory and at the Max-Planck-Institut für Radioastronomie, each handling one of the two frequency bands. Differences between the two independent correlators were shown to be negligible through the exchange of a few identical scans for cross comparison. At correlation, signals were aligned to a common time reference using an a priori Earth geometry and clock model.

A subsequent fringe-fitting step identified detections in correlated signal power while phase calibrating the data for residual delays and atmospheric effects. Using ALMA as a highly sensitive reference station enabled critical corrections for ionospheric and tropospheric distortions at the other sites. Fringe fitting was performed with three independent automated pipelines, each tailored to the specific characteristics of the EHT observations, such as the wide bandwidth, susceptibility to atmospheric turbulence, and array heterogeneity (Blackburn et al. 2019; Janssen et al. 2019, Paper III). The pipelines made use of standard software for the processing of radio-interferometric data (Greisen 2003; Whitney et al. 2004; McMullin et al. 2007, I. M. van Bemmel et al. 2019, in preparation).

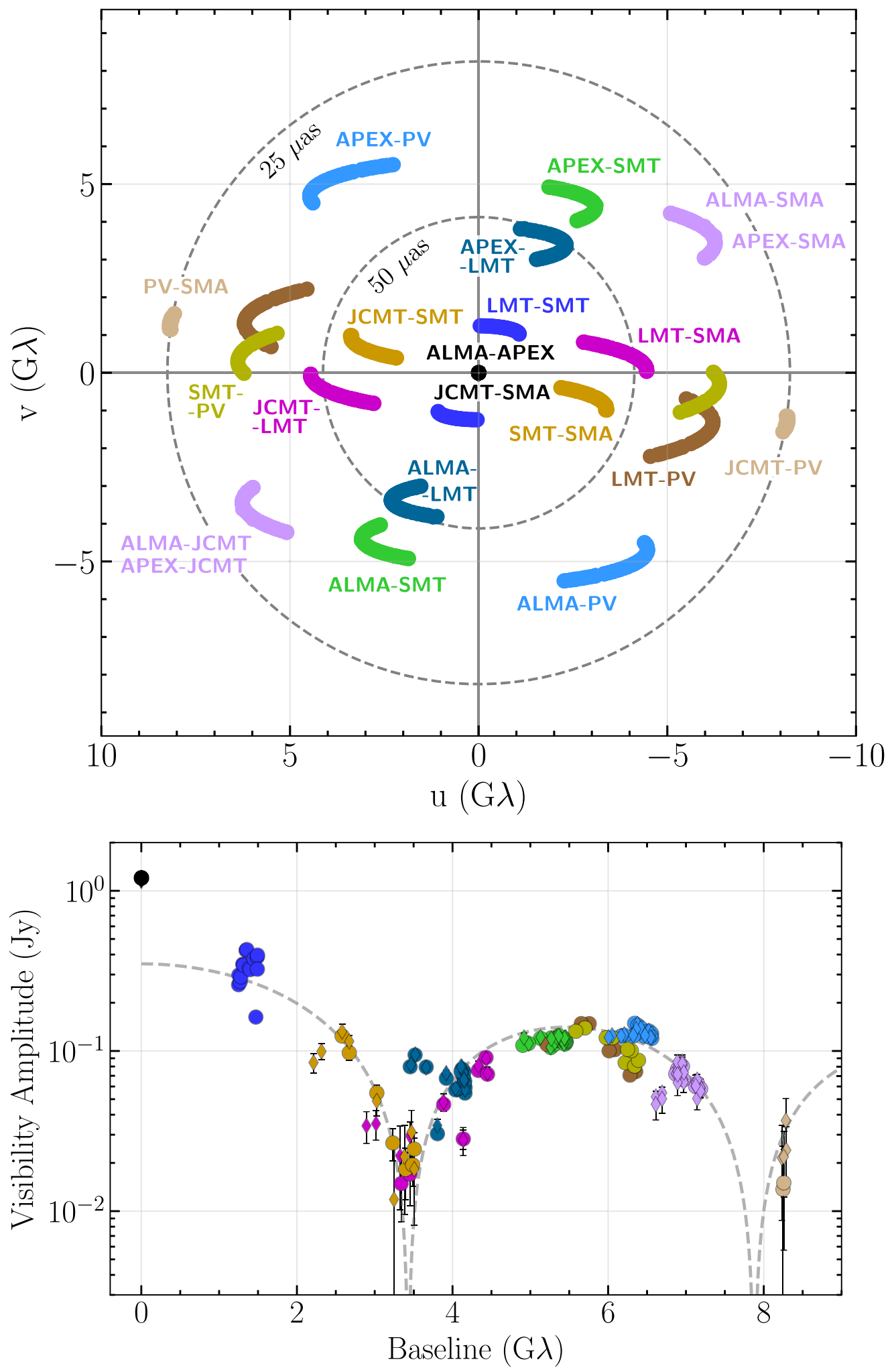

**Figure 2.** Top: ($u$, $v$) coverage for M87*, aggregated over all four days of the observations. ($u$, $v$) coordinates for each antenna pair are the source-projected baseline length in units of the observing wavelength $\lambda$ and are given for conjugate pairs. Baselines to ALMA/APEX and to JCMT/SMA are redundant. Dotted circular lines indicate baseline lengths corresponding to fringe spacings of 50 and 25 $\mu$as. Bottom: final calibrated visibility amplitudes of M87* as a function of projected baseline length on April 11. Redundant baselines to APEX and JCMT are plotted as diamonds. Error bars correspond to thermal (statistical) uncertainties. The Fourier transform of an azimuthally symmetric thin ring model with diameter 46 $\mu$as is also shown with a dashed line for comparison.

Data from the fringe-fitting pipelines were scaled from correlation coefficients to a uniform physical flux density scale (in Jansky) by using an independent a priori estimate of the sensitivity of each telescope. The accuracies of the derived station sensitivities were estimated to be 5%–10% in amplitude, although certain uncharacterized losses (e.g., from poor pointing or focus) can exceed the error budget. By assuming total flux density values derived from ALMA interferometric data (Goddi et al. 2019) and utilizing array redundancy via network calibration (Paper III), we refined the absolute amplitude calibration of telescopes that are colocated and have redundant baselines, i.e., ALMA/APEX and JCMT/SMA.

The median scan-averaged signal-to-noise ratio for M87* was >10 on non-ALMA baselines and >100 on baselines to ALMA, leading to small statistical errors in visibility amplitude and phase. Comparisons between the three independent pipelines, the two polarizations, and the two frequency bands enabled estimation of systematic baseline errors of around 1° in visibility phase and 2% for visibility amplitudes. These small limiting errors remain after fitting station sensitivities and unknown station phases via self-calibration (Pearson & Readhead 1984) and affect interferometric closure quantities (Rogers et al. 1974; Readhead et al. 1980). Following data validation and pipeline comparisons, a single pipeline output was designated as the primary data set of the first EHT science data release and used for subsequent results, while the outputs of the other two pipelines offer supporting validation data sets.

The final calibrated complex visibilities $V(u, v)$ correspond to the Fourier components of the brightness distribution on the sky at spatial frequency ($u$, $v$) determined by the projected baseline expressed in units of the observing wavelength (van Cittert 1934; Thompson et al. 2017). Figure 2 shows the ($u$, $v$) coverage and calibrated visibility amplitudes of M87* for April 11. The visibility amplitudes resemble those of a thin ring (i.e., a Bessel function $J_0$; see Figure 10.12 in Thompson et al. 2017). Such a ring model with diameter 46 $\mu$as has a first null at 3.4 G$\lambda$, matching the minimum in observed flux density and is consistent with a reduced flux density on the longest Hawai'i–Spain baseline (JCMT/SMA-PV) near 8 G$\lambda$. This particular ring model, shown with a dashed line in the bottom panel of Figure 2, is only illustrative and does not fit all features in the data. First, visibility amplitudes on the shortest VLBI baselines suggest that about half of the compact flux density seen on the ~2 km ALMA–APEX baseline is resolved out by the interferometer beam (Paper IV). Second, differences in the depth of the first minimum as a function of orientation, as well as highly nonzero measured closure phases, indicate some degree of asymmetry in the source (Papers III, VI). Finally, the visibility amplitudes represent only half of the information available to us. We will next explore images and more complex geometrical models that can fit the measured visibility amplitudes and phases.

## 5. Images and Features

We reconstructed images from the calibrated EHT visibilities, which provide results that are independent of models (Paper IV). However, there are two major challenges in reconstructing images from EHT data. First, EHT baselines sample a limited range of spatial frequencies, corresponding to angular scales between 25 and 160 $\mu$as. Because the ($u$, $v$) plane is only sparsely sampled (Figure 2), the inverse problem is under-constrained. Second, the measured visibilities lack absolute phase calibration and can have large amplitude calibration uncertainties.

To address these challenges, imaging algorithms incorporate additional assumptions and constraints that are designed to produce images that are physically plausible (e.g., positive and compact) or conservative (e.g., smooth), while remaining consistent with the data. We explored two classes of algorithms for reconstructing images from EHT data. The first class of algorithms is the traditional `CLEAN` approach used in radio interferometry (e.g., Högbom 1974; Clark 1980). `CLEAN` is an inverse-modeling approach that deconvolves the interferometer point-spread function from the Fourier-transformed visibilities. When applying `CLEAN`, it is necessary to iteratively self-calibrate the data between rounds of imaging to solve for time-variable phase and amplitude errors in the data. The second class of algorithms is the so-called regularized maximum likelihood (RML; e.g., Narayan & Nityananda 1986; Wiaux et al. 2009; Thiébaut 2013). RML is a forward-modeling approach that searches for an image that is not only consistent with the observed data but also favors specified image properties (e.g., smoothness or compactness). As with CLEAN, RML methods typically iterate between imaging and self-calibration, although they can also be used to image directly on robust closure quantities immune to station-based calibration errors. RML methods have been extensively developed for the EHT (e.g., Honma et al. 2014; Bouman et al. 2016; Akiyama et al. 2017; Chael et al. 2018b; see also Paper IV).

Every imaging algorithm has a variety of free parameters that can significantly affect the final image. We adopted a two-stage imaging approach to control and evaluate biases in the reconstructions from our choices of these parameters. In the first stage, four teams worked independently to reconstruct the first EHT images of M87* using an early engineering data release. The teams worked without interaction to minimize shared bias, yet each produced an image with a similar prominent feature: a ring of diameter ~38–44 $\mu$as with enhanced brightness to the south (see Figure 4 in Paper IV).

In the second imaging stage, we developed three imaging pipelines, each using a different software package and associated methodology. Each pipeline surveyed a range of imaging parameters, producing between $\sim 10^3$ and $10^4$ images from different parameter combinations. We determined a "Top-Set" of parameter combinations that both produced images of M87* that were consistent with the observed data and that reconstructed accurate images from synthetic data sets corresponding to four known geometric models (ring, crescent, filled disk, and asymmetric double source). For all pipelines, the Top-Set images showed an asymmetric ring with a diameter of ~40 $\mu$as, with differences arising primarily in the effective angular resolutions achieved by different methods.

For each pipeline, we determined the single combination of fiducial imaging parameters out of the Top-Set that performed best across all the synthetic data sets and for each associated imaging methodology (see Figure 11 in Paper IV). Because the angular resolutions of the reconstructed images vary among the pipelines, we blurred each image with a circular Gaussian to a common, conservative angular resolution of 20 $\mu$as. The top part of Figure 3 shows an image of M87* on April 11 obtained by averaging the three pipelines' blurred fiducial images. The image is dominated by a ring with an asymmetric azimuthal profile that is oriented at a position angle ~170° east of north. Although the measured position angle increases by ~20° between the first two days and the last two days, the image features are broadly consistent across the different imaging methods and across all four observing days. This is shown in the bottom part of Figure 3, which reports the images on different days (see also Figure 15 in Paper IV). These results are also consistent with those obtained from visibility-domain fitting of geometric and general-relativistic magnetohydrodynamics (GRMHD) models (Paper VI).

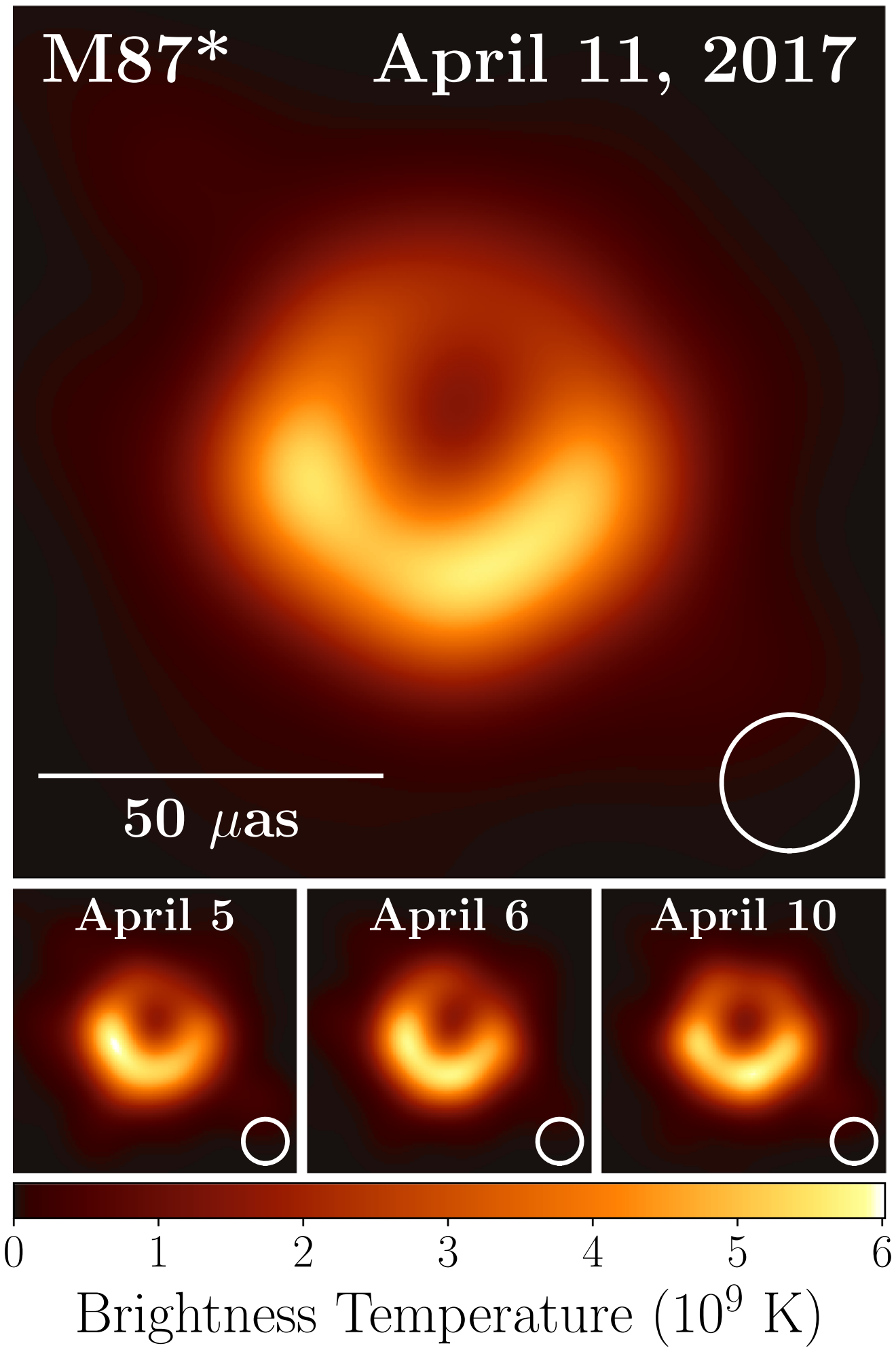

**Figure 3.** Top: EHT image of M87* from observations on 2017 April 11 as a representative example of the images collected in the 2017 campaign. The image is the average of three different imaging methods after convolving each with a circular Gaussian kernel to give matched resolutions. The largest of the three kernels (20 $\mu$as FWHM) is shown in the lower right. The image is shown in units of brightness temperature, $T_b = S\lambda^2/2k_B\Omega$, where $S$ is the flux density, $\lambda$ is the observing wavelength, $k_B$ is the Boltzmann constant, and $\Omega$ is the solid angle of the resolution element. Bottom: similar images taken over different days showing the stability of the basic image structure and the equivalence among different days. North is up and east is to the left.

## 6. Theoretical Modeling

The appearance of M87* has been modeled successfully using GRMHD simulations, which describe a turbulent, hot, magnetized disk orbiting a Kerr black hole. They naturally produce a powerful jet and can explain the broadband spectral energy distribution observed in LLAGNs. At a wavelength of 1.3 mm, and as observed here, the simulations also predict a shadow and an asymmetric emission ring. The latter does not necessarily coincide with the innermost stable circular orbit, or ISCO, and is instead related to the lensed photon ring. To explore this scenario in great detail, we have built a library of synthetic images (Image Library) describing magnetized accretion flows onto black holes in GR[145] (Paper V). The images themselves are produced from a library of simulations (Simulation Library) collecting the results of four codes solving the equations of GRMHD (Gammie et al. 2003; Sądowski et al. 2014; Porth et al. 2017; Liska et al. 2018). The elements of the Simulation Library have been coupled to three different general-relativistic ray-tracing and radiative-transfer codes (GRRT, Bronzwaer et al. 2018; Mościbrodzka & Gammie 2018; Z. Younsi et al. 2019, in preparation). We limit ourselves to providing here a brief description of the initial setups and the physical scenarios explored in the simulations; see Paper V for details on both the GRMHD and GRRT codes, which have been cross-validated

[145] More exotic spacetimes, such as dilaton black holes, boson stars, and gravastars, have also been considered (Paper V).

for accuracy and consistency (Gold et al. 2019; Porth et al. 2019).

A typical GRMHD simulation in the library is characterized by two parameters: the dimensionless spin $a_* \equiv Jc/GM^2$, where $J$ and $M$ are, respectively, the spin angular momentum and mass of the black hole, and the net dimensionless magnetic flux over the event horizon $\phi \equiv \Phi/(\dot{M}R_g^2)^{1/2}$, where $\Phi$ and $\dot{M}$ are the magnetic flux and mass flux (or accretion rate) across the horizon, respectively. Since the GRMHD simulations scale with the black hole mass, $M$ is set only at the time of producing the synthetic images with the GRRT codes. The magnetic flux is generally nonzero because magnetic field is trapped in the black hole by the accretion flow and sustained by currents in the surrounding plasma.

These two parameters allow us to describe accretion disks that are either prograde ($a_* \geqslant 0$) or retrograde ($a_* < 0$) with respect to the black hole spin axis, and whose accretion flows are either "SANE" (from "Standard and Normal Evolution," Narayan et al. 2012) with $\phi \sim 1$, or "MAD" (from "Magnetically Arrested Disk," Narayan et al. 2003) with $\phi \sim 15$.[146] In essence, SANE accretion flows are characterized by moderate dimensionless magnetic flux and result from initial magnetic fields that are smaller than those in MAD flows. Furthermore, the opening angles of the magnetic funnel in SANE flows are generically smaller than those in MAD flows. Varying $a_*$ and $\phi$, we have performed 43 high-resolution, three-dimensional and long-term simulations covering well the physical properties of magnetized accretion flows onto Kerr black holes.

All GRMHD simulations were initialized with a weakly magnetized torus orbiting around the black hole and driven into a turbulent state by instabilities, including the magnetorotational instability (Balbus & Hawley 1991), rapidly reaching a quasi-stationary state. Once a simulation was completed, the relevant flow properties at different times are collected to be employed for the further post-processing of the GRRT codes. The generation of synthetic images requires, besides the properties of the fluid (magnetic field, velocity field, and rest-mass density), also the emission and absorption coefficients, the inclination $i$ (the angle between the accretion-flow angular-momentum vector and the line of sight), the position angle PA (the angle east of north, i.e., counter-clockwise on our images, of the projection on the sky of the accretion-flow angular momentum), the black hole mass $M$ and distance $D$ to the observer.

Because the photons at 1.3 mm wavelength observed by the EHT are believed to be produced by synchrotron emission, whose absorption and emission coefficients depend on the electron distribution function, we consider the plasma to be composed of electrons and ions that have the same temperature in the magnetically dominated regions of the flow (funnel), but have a substantially different temperature in the gas dominated regions (disk midplane). In particular, we consider the plasma to be composed of nonrelativistic ions with temperature $T_i$ and relativistic electrons with temperature $T_e$. A simple prescription for the ratio of the temperatures of the two species can then be imposed in terms of a single parameter (Mościbrodzka et al. 2016), such that the bulk of the emission comes either from weakly magnetized (small $R_{\rm high}$, $Te \simeq Ti/R_{\rm high}$) or strongly magnetized (large $R_{\rm high}$, $Te \simeq Ti$) regions. In SANE models, the disk (jet) is weakly (strongly) magnetized, so that low (high) $R_{\rm high}$ models produce most of the emission in the disk (jet). In MAD models, there are strongly magnetized regions everywhere and the emission is mostly from the disk midplane. While this prescription is not the only one possible, it has the advantage of being simple, sufficiently generic, and robust (see Paper V for a discussion of nonthermal particles and radiative cooling).

Since each GRMHD simulation can be used to describe several different physical scenarios by changing the prescribed electron distribution function, we have used the Simulation Library to generate more than 420 different physical scenarios. Each scenario is then used to generate hundreds of snapshots at different times in the simulation leading to more than 62,000 objects in the Image Library. From the images we have created model visibilities that correspond to the EHT observing schedule and compared them to the measured VLBI visibilities as detailed in Paper VI.

As an example, the top row of Figure 4 shows three GRMHD model snapshots from the Image Library with different spins and flow type, which fitted closure phases and amplitudes of the April 11 data best. For these models we produced simulated visibilities for the April 11 schedule and weather parameters and calibrated them with a synthetic data generation and calibration pipeline (Blecher et al. 2017; Janssen et al. 2019; Roelofs et al. 2019a). The simulated data were then imaged with the same pipeline used for the observed images. The similarities between the simulated images (bottom row of Figure 4) and the observed images (Figure 3) are remarkable.

Overall, when combining all the information contained in both the Simulation Library and Image Library, the physical origin of the emission features of the image observed in M87* can be summarized as follows.

First, the observed image is consistent with the hypothesis that it is produced by a magnetized accretion flow orbiting within a few $r_g$ of the event horizon of a Kerr black hole. The asymmetric ring is produced by a combination of strong gravitational lensing and relativistic beaming, while the central flux depression is the observational signature of the black hole shadow. Interestingly, all of the accretion models are consistent with the EHT image, except for the $a_* = -0.94$ MAD models, which fail to produce images that are sufficiently stable (i.e., the variance among snapshots is too large to be statistically consistent with the observed image).

Second, the north–south asymmetry in the emission ring is controlled by the black hole spin and can be used to deduce its orientation. In corotating disk models (where the angular momentum of the matter and of the black hole are aligned), the funnel wall, or jet sheath, rotates with the disk and the black hole; in counterrotating disk models, instead, the luminous funnel wall rotates with the black hole but against the disk. The north–south asymmetry is consistent with models in which the black hole spin is pointing away from Earth and inconsistent with models in which the spin points toward Earth.

Third, adopting an inclination of 17° between the approaching jet and the line of sight (Walker et al. 2018), the west orientation of the jet, and a corotating disk model, matter in the bottom part of the image is moving toward the observer (clockwise rotation as seen from Earth). This is consistent with the rotation of the ionized gas on scales of 20 pc, i.e., 7000 $r_g$ (Ford et al. 1994; Walsh et al. 2013) and with the inferred sense of rotation from VLBI observations at 7 mm (Walker et al. 2018).

Finally, models with $a_* = 0$ are disfavored by the very conservative observational requirement that the jet power be $P_{\rm jet} > 10^{42}\,{\rm erg\,s^{-1}}$. Furthermore, in those models that produce a sufficiently powerful jet, it is powered by extraction of black hole

[146] We here use Heaviside units, where a factor of $\sqrt{4\pi}$ is absorbed into the definition of the magnetic field.

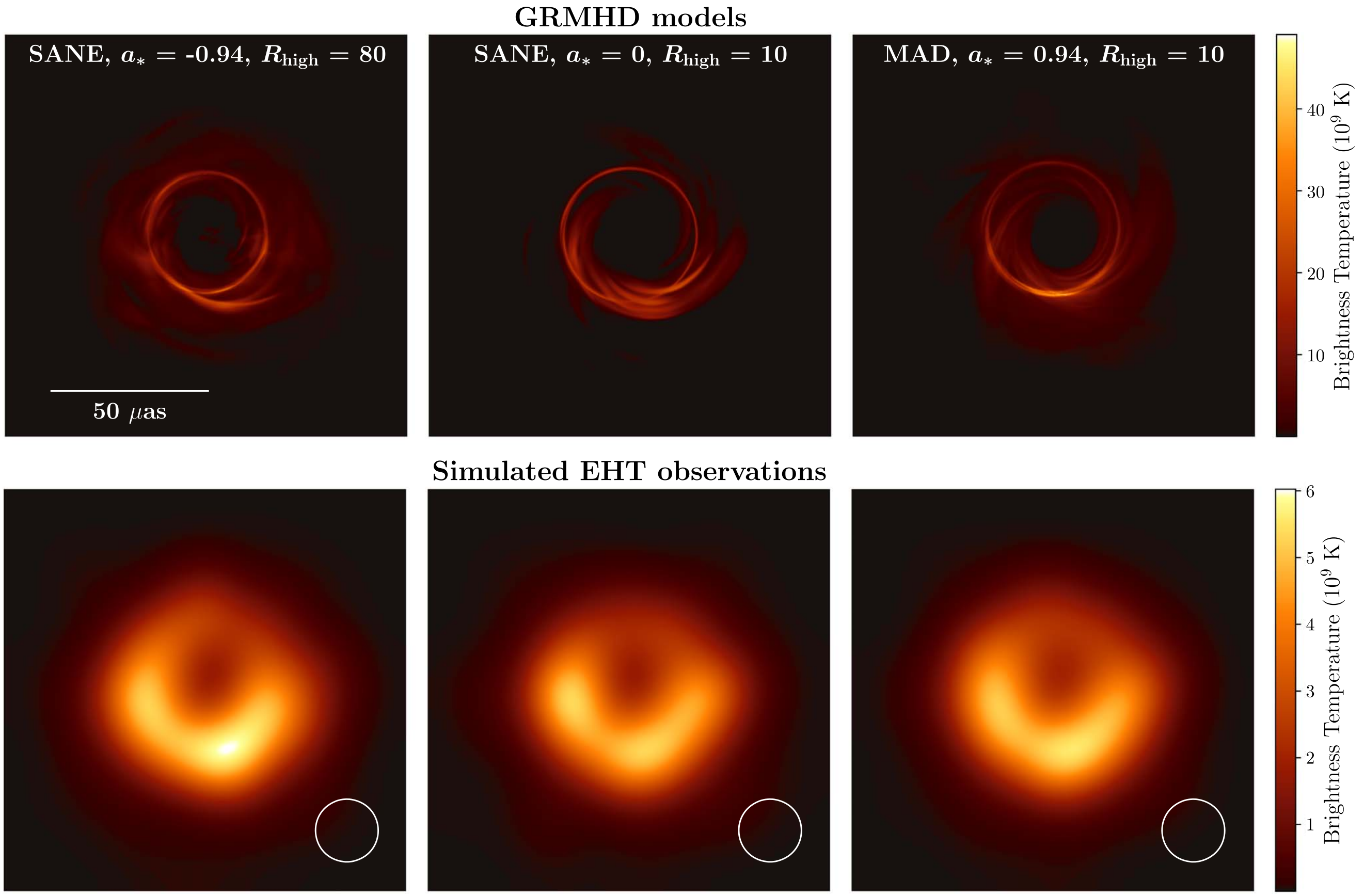

**Figure 4.** Top: three example models of some of the best-fitting snapshots from the image library of GRMHD simulations for April 11 corresponding to different spin parameters and accretion flows. Bottom: the same theoretical models, processed through a VLBI simulation pipeline with the same schedule, telescope characteristics, and weather parameters as in the April 11 run and imaged in the same way as Figure 3. Note that although the fit to the observations is equally good in the three cases, they refer to radically different physical scenarios; this highlights that a single good fit does not imply that a model is preferred over others (see Paper V).

spin energy through mechanisms akin to the Blandford–Znajek process.

## 7. Model Comparison and Parameter Estimation

In Paper VI, the black hole mass is derived from fitting to the visibility data of geometric and GRMHD models, as well as from measurements of the ring diameter in the image domain. Our measurements remain consistent across methodologies, algorithms, data representations, and observed data sets.

Motivated by the asymmetric emission ring structures seen in the reconstructed images (Section 5) and the similar emission structures seen in the images from GRMHD simulations (Section 6), we developed a family of geometric crescent models (see, e.g., Kamruddin & Dexter 2013) to compare directly to the visibility data. We used two distinct Bayesian-inference algorithms and demonstrate that such crescent models are statistically preferred over other comparably complex geometric models that we have explored. We find that the crescent models provide fits to the data that are statistically comparable to those of the reconstructed images presented in Section 5, allowing us to determine the basic parameters of the crescents. The best-fit models for the asymmetric emission ring have diameters of $43 \pm 0.9\,\mu$as and fractional widths relative to the diameter of $<0.5$. The emission drops sharply interior to the asymmetric emission ring with the central depression having a brightness $<10\%$ of the average brightness in the ring.

The diameters of the geometric crescent models measure the characteristic sizes of the emitting regions that surround the shadows and not the sizes of the shadows themselves (see, e.g., Psaltis et al. 2015; Johannsen et al. 2016; Kuramochi et al. 2018, for potential biases).

We model the crescent angular diameter $d$ in terms of the gravitational radius and distance, $\theta_{\rm g} \equiv GM/c^2D$, as $d = \alpha\theta_{\rm g}$, where $\alpha$ is a function of spin, inclination, and $R_{\rm high}$ ($\alpha \simeq 9.6$–$10.4$ corresponds to emission from the lensed photon ring only). We calibrate $\alpha$ by fitting the geometric crescent models to a large number of visibility data generated from the Image Library. We can also fit the model visibilities generated from the Image Library to the M87* data, which allows us to measure $\theta_{\rm g}$ directly. However, such a procedure is complicated by the stochastic nature of the emission in the accretion flow (see, e.g., Kim et al. 2016). To account for this turbulent structure, we developed a formalism and multiple algorithms that estimate the statistics of the stochastic components using ensembles of images from individual GRMHD simulations. We find that the visibility data are not inconsistent with being a realization of many of the GRMHD simulations. We conclude that the recovered model parameters are consistent across algorithms.

Finally, we extract ring diameter, width, and shape directly from reconstructed images (see Section 5). The results are consistent with the parameter estimates from geometric crescent models. Following the same GRMHD calibration

**Table 1**
Parameters of M87*

| Parameter | Estimate |
|---|---|
| Ring diameter [a] $d$ | $42 \pm 3$ $\mu$as |
| Ring width [a] | $<20$ $\mu$as |
| Crescent contrast [b] | $>10:1$ |
| Axial ratio [a] | $<4:3$ |
| Orientation PA | 150°–200° east of north |
| $\theta_g = GM/Dc^2$ [c] | $3.8 \pm 0.4$ $\mu$as |
| $\alpha = d/\theta_g$ [d] | $11^{+0.5}_{-0.3}$ |
| $M$[c] | $(6.5 \pm 0.7) \times 10^9\ M_\odot$ |
| **Parameter** | **Prior Estimate** |
| $D$ [e] | $(16.8 \pm 0.8)$ Mpc |
| $M$(stars) [e] | $6.2^{+1.1}_{-0.6} \times 10^9\ M_\odot$ |
| $M$(gas) [e] | $3.5^{+0.9}_{-0.3} \times 10^9\ M_\odot$ |

**Notes.**
[a] Derived from the image domain.
[b] Derived from crescent model fitting.
[c] The mass and systematic errors are averages of the three methods (geometric models, GRMHD models, and image domain ring extraction).
[d] The exact value depends on the method used to extract $d$, which is reflected in the range given.
[e] Rederived from likelihood distributions (Paper VI).

procedure, we infer values of $\theta_g$ and $\alpha$ for regularized maximum likelihood and CLEAN reconstructed images.

Combining results from all methods, we measure emission region diameters of $42 \pm 3$ $\mu$as, angular sizes of the gravitational radius $\theta_g = 3.8 \pm 0.4$ $\mu$as, and scaling factors in the range $\alpha = 10.7$–$11.5$, with associated errors of ~10%. For the distance of $16.8 \pm 0.8$ Mpc adopted here, the black hole mass is $M = (6.5 \pm 0.7) \times 10^9\ M_\odot$; the systematic error refers to the 68% confidence level and is much larger than the statistical error of $0.2 \times 10^9\ M_\odot$. Moreover, by tracing the peak of the emission in the ring we can determine the shape of the image and obtain a ratio between major and minor axis of the ring that is $\lesssim 4:3$; this corresponds to a $\lesssim 10\%$ deviation from circularity in terms of root-mean-square distance from an average radius.

Table 1 summarizes the measured parameters of the image features and the inferred black hole properties based on data from all bands and all days combined. The inferred black hole mass strongly favors the measurement based on stellar dynamics (Gebhardt et al. 2011). The size, asymmetry, brightness contrast, and circularity of the reconstructed images and geometric models, as well as the success of the GRMHD simulations in describing the interferometric data, are consistent with the EHT images of M87* being associated with strongly lensed emission from the vicinity of a Kerr black hole.

## 8. Discussion

A number of elements reinforce the robustness of our image and the conclusion that it is consistent with the shadow of a black hole as predicted by GR. First, our analysis has used multiple independent calibration and imaging techniques, as well as four independent data sets taken on four different days in two separate frequency bands. Second, the image structure matches previous predictions well (Dexter et al. 2012; Mościbrodzka et al. 2016) and is well reproduced by our extensive modeling effort presented in Section 6. Third, because our measurement of the black hole mass in M87* is not inconsistent with all of the prior mass measurements, this allows us to conclude that the null hypothesis of the Kerr metric (Psaltis et al. 2015; Johannsen et al. 2016), namely, the assumption that the black hole is described by the Kerr metric, has not been violated. Fourth, the observed emission ring reconstructed in our images is close to circular with an axial ratio $\lesssim 4:3$; similarly, the time average images from our GRMHD simulations also show a circular shape. After associating to the shape of the shadow a deviation from the circularity—measured in terms of root-mean-square distance from an average radius in the image—that is $\lesssim 10\%$, we can set an initial limit of order four on relative deviations of the quadrupole moment from the Kerr value (Johannsen & Psaltis 2010). Stated differently, if $Q$ is the quadrupole moment of a Kerr black hole and $\Delta Q$ the deviation as deduced from circularity, our measurement—and the fact that the inclination angle is assumed to be small—implies that $\Delta Q/Q \lesssim 4$ ($\Delta Q/Q = \varepsilon$ in Johannsen & Psaltis 2010).

Finally, when comparing the visibility amplitudes of M87* with 2009 and 2012 data (Doeleman et al. 2012; Akiyama et al. 2015), the overall radio core size at a wavelength of 1.3 mm has not changed appreciably, despite variability in total flux density. This stability is consistent with the expectation that the size of the shadow is a feature tied to the mass of the black hole and not to properties of a variable plasma flow.

It is also straightforward to reject some alternative astrophysical interpretations. For instance, the image is unlikely to be produced by a jet-feature as multi-epoch VLBI observations of the plasma jet in M87 (Walker et al. 2018) on scales outside the horizon do not show circular rings. The same is typically true for AGN jets in large VLBI surveys (Lister et al. 2018). Similarly, were the apparent ring a random alignment of emission blobs, they should also have moved away at relativistic speeds, i.e., at ~5 $\mu$as day$^{-1}$ (Kim et al. 2018b), leading to measurable structural changes and sizes. GRMHD models of hollow jet cones could show under extreme conditions stable ring features (Pu et al. 2017), but this effect is included to a certain extent in our Simulation Library for models with $R_{\rm high} > 10$. Finally, an Einstein ring formed by gravitational lensing of a bright region in the counter-jet would require a fine-tuned alignment and a size larger than that measured in 2012 and 2009.

At the same time, it is more difficult to rule out alternatives to black holes in GR, because a shadow can be produced by any compact object with a spacetime characterized by unstable circular photon orbits (Mizuno et al. 2018). Indeed, while the Kerr metric remains a solution in some alternative theories of gravity (Barausse & Sotiriou 2008; Psaltis et al. 2008), non-Kerr black hole solutions do exist in a variety of such modified theories (Berti et al. 2015). Furthermore, exotic alternatives to black holes, such as naked singularities (Shaikh et al. 2019), boson stars (Kaup 1968; Liebling & Palenzuela 2012), and gravastars (Mazur & Mottola 2004; Chirenti & Rezzolla 2007), are admissible solutions within GR and provide concrete, albeit contrived, models. Some of such exotic compact objects can already be shown to be incompatible with our observations given our maximum mass prior. For example, the shadows of naked singularities associated with Kerr spacetimes with $|a_*| > 1$ are substantially smaller and very asymmetric compared to those of Kerr black holes (Bambi & Freese 2009). Also, some commonly used types of wormholes (Bambi 2013) predict much smaller shadows than we have measured.

However, other compact-object candidates need to be analyzed with more care. Boson stars are an example of compact objects having circular photon orbits but without a surface or an event horizon. In such a spacetime, null geodesics are redirected outwards toward distant observers (Cunha et al. 2016), so that the shadow can in principle be filled with emission from lensed images of distant radio sources generating a complex mirror image of the sky. More importantly, accretion flows onto boson stars behave differently as they do not produce jets but stalled accretion tori that make them distinguishable from black holes (Olivares et al. 2019). Gravastars provide examples of compact objects having unstable photon orbits and a hard surface, but not an event horizon. In this case, while a single image of the accretion flow could in principle be very similar to that coming from a black hole, differences of the flow dynamics at the stellar surface (H. Olivares et al. 2019, in preparation), strong magnetic fields (Lobanov 2017), or excess radiation in the infrared (Broderick & Narayan 2006) would allow one to distinguish a gravastar from a black hole.

Altogether, the results derived here provide a new way to study compact-object spacetimes and are complementary to the detection of gravitational waves from coalescing stellar-mass black holes with LIGO/Virgo (Abbott et al. 2016). Our constraints on deviations from the Kerr geometry rely only on the validity of the equivalence principle and are agnostic about the underlying theory of gravity, but can be used to measure, with ever improved precision, the parameters of the background metric. On the other hand, current gravitational-wave observations of mergers probe the dynamics of the underlying theory, but cannot rely on the possibility of multiple and repeated measurements of the same source.

To underline the complementarity of gravitational-wave and electromagnetic observations of black holes, we note that a basic feature of black holes in GR is that their size scales linearly with mass. Combining our constraints on the supermassive black hole in M87 with those on the stellar-mass black holes detected via gravitational waves we can infer that this property holds over eight orders of magnitude. While this invariance is a basic prediction of GR, which is a scale-free theory, it need not be satisfied in other theories that introduce a scale to the gravitational field.

Finally, the radio core in M87 is quite typical for powerful radio jets in general. It falls on the fundamental plane of black hole activity for radio cores (Falcke et al. 2004), connecting via simple scaling laws the radio and X-ray properties of low-luminosity black hole candidates across vastly different mass and accretion rate scales. This suggests that they are powered by a scale-invariant common object. Therefore, establishing the black hole nature for M87* also supports the general paradigm that black holes are the power source for active galaxies.

## 9. Conclusion and Outlook

We have assembled the EHT, a global VLBI array operating at a wavelength of 1.3 mm and imaged horizon-scale structures around the supermassive black hole candidate in M87. Using multiple independent calibration, imaging, and analysis methods, we find the image to be dominated by a ring structure of $42 \pm 3$ $\mu$as diameter that is brighter in the south. This structure has a central brightness depression with a contrast of >10:1, which we identify with the black hole shadow. Comparing the data with an extensive library of synthetic images obtained from GRMHD simulations covering different physical scenarios and plasma conditions reveals that the basic features of our image are relatively independent of the detailed astrophysical model. This allows us to derive an estimate for the black hole mass of $M = (6.5 \pm 0.7) \times 10^9\, M_\odot$. Based on our modeling and information on the inclination angle, we derive the sense of rotation of the black hole to be in the clockwise direction, i.e., the spin of the black hole points away from us. The brightness excess in the south part of the emission ring is explained as relativistic beaming of material rotating in the clockwise direction as seen by the observer, i.e., the bottom part of the emission region is moving toward the observer.

Future observations and further analysis will test the stability, shape, and depth of the shadow more accurately. One of its key features is that it should remain largely constant with time as the mass of M87* is not expected to change measurably on human timescales. Polarimetric analysis of the images, which we will report in the future, will provide information on the accretion rate via Faraday rotation (Bower et al. 2003; Marrone et al. 2007; Kuo et al. 2014; Mościbrodzka et al. 2017) and on the magnetic flux. Higher-resolution images can be achieved by going to a shorter wavelength, i.e., 0.8 mm (345 GHz), by adding more telescopes and, in a more distant future, with space-based interferometry (Kardashev et al. 2014; Fish et al. 2019; Palumbo et al. 2019; F. Roelofs et al. 2019b, in preparation).

Another primary EHT source, Sgr A*, has a precisely measured mass three orders of magnitude smaller than that of M87*, with dynamical timescales of minutes instead of days. Observing the shadow of Sgr A* will require accounting for this variability and mitigation of scattering effects caused by the interstellar medium (Johnson 2016; Lu et al. 2016; Bouman et al. 2018). Time dependent nonimaging analysis can be used to potentially track the motion of emitting matter near the black hole, as reported recently through interferometric observations in the near-infrared (Gravity Collaboration et al. 2018b). Such observations provide separate tests and probes of GR on yet another mass scale (Broderick & Loeb 2005; Doeleman et al. 2009b; Roelofs et al. 2017; Medeiros et al. 2017).

In conclusion, we have shown that direct studies of the event horizon shadow of supermassive black hole candidates are now possible via electromagnetic waves, thus transforming this elusive boundary from a mathematical concept to a physical entity that can be studied and tested via repeated astronomical observations.

The authors of this Letter thank the following organizations and programs: the Academy of Finland (projects 274477, 284495, 312496); the Advanced European Network of E-infrastructures for Astronomy with the SKA (AENEAS) project, supported by the European Commission Framework Programme Horizon 2020 Research and Innovation action under grant agreement 731016; the Alexander von Humboldt Stiftung; the Black Hole Initiative at Harvard University, through a grant (60477) from the John Templeton Foundation; the China Scholarship Council; Comisión Nacional de Investigación Científica y Tecnológica (CONICYT, Chile, via PIA ACT172033, Fondecyt 1171506, BASAL AFB-170002, ALMA-conicyt 31140007); Consejo Nacional de Ciencia y Tecnología (CONACYT, Mexico, projects 104497, 275201, 279006, 281692); the Delaney Family via the Delaney Family John A. Wheeler Chair at Perimeter Institute; Dirección General de Asuntos del Personal Académico-Universidad Nacional

Autónoma de México (DGAPA-UNAM, project IN112417); the European Research Council (ERC) Synergy Grant "BlackHoleCam: Imaging the Event Horizon of Black Holes" (grant 610058); the Generalitat Valenciana postdoctoral grant APOSTD/2018/177; the Gordon and Betty Moore Foundation (grants GBMF-3561, GBMF-5278); the Istituto Nazionale di Fisica Nucleare (INFN) sezione di Napoli, iniziative specifiche TEONGRAV; the International Max Planck Research School for Astronomy and Astrophysics at the Universities of Bonn and Cologne; the Jansky Fellowship program of the National Radio Astronomy Observatory (NRAO); the Japanese Government (Monbukagakusho: MEXT) Scholarship; the Japan Society for the Promotion of Science (JSPS) Grant-in-Aid for JSPS Research Fellowship (JP17J08829); JSPS Overseas Research Fellowships; the Key Research Program of Frontier Sciences, Chinese Academy of Sciences (CAS, grants QYZDJ-SSW-SLH057, QYZDJ-SSW-SYS008); the Leverhulme Trust Early Career Research Fellowship; the Max-Planck-Gesellschaft (MPG); the Max Planck Partner Group of the MPG and the CAS; the MEXT/JSPS KAKENHI (grants 18KK0090, JP18K13594, JP18K03656, JP18H03721, 18K03709, 18H01245, 25120007); the MIT International Science and Technology Initiatives (MISTI) Funds; the Ministry of Science and Technology (MOST) of Taiwan (105-2112-M-001-025-MY3, 106-2112-M-001-011, 106-2119-M-001-027, 107-2119-M-001-017, 107-2119-M-001-020, and 107-2119-M-110-005); the National Aeronautics and Space Administration (NASA, Fermi Guest Investigator grant 80NSSC17K0649); the National Institute of Natural Sciences (NINS) of Japan; the National Key Research and Development Program of China (grant 2016YFA0400704, 2016YFA0400702); the National Science Foundation (NSF, grants AST-0096454, AST-0352953, AST-0521233, AST-0705062, AST-0905844, AST-0922984, AST-1126433, AST-1140030, DGE-1144085, AST-1207704, AST-1207730, AST-1207752, MRI-1228509, OPP-1248097, AST-1310896, AST-1312651, AST-1337663, AST-1440254, AST-1555365, AST-1715061, AST-1615796, AST-1614868, AST-1716327, OISE-1743747, AST-1816420); the Natural Science Foundation of China (grants 11573051, 11633006, 11650110427, 10625314, 11721303, 11725312, 11873028, 11873073, U1531245, 11473010); the Natural Sciences and Engineering Research Council of Canada (NSERC, including a Discovery Grant and the NSERC Alexander Graham Bell Canada Graduate Scholarships-Doctoral Program); the National Youth Thousand Talents Program of China; the National Research Foundation of Korea (grant 2015-R1D1A1A01056807, the Global PhD Fellowship Grant: NRF-2015H1A2A1033752, and the Korea Research Fellowship Program: NRF-2015H1D3A1066561); the Netherlands Organization for Scientific Research (NWO) VICI award (grant 639.043.513) and Spinoza Prize (SPI 78-409); the New Scientific Frontiers with Precision Radio Interferometry Fellowship awarded by the South African Radio Astronomy Observatory (SARAO), which is a facility of the National Research Foundation (NRF), an agency of the Department of Science and Technology (DST) of South Africa; the Onsala Space Observatory (OSO) national infrastructure, for the provisioning of its facilities/observational support (OSO receives funding through the Swedish Research Council under grant 2017-00648); the Perimeter Institute for Theoretical Physics (research at Perimeter Institute is supported by the Government of Canada through the Department of Innovation, Science and Economic Development Canada and by the Province of Ontario through the Ministry of Economic Development, Job Creation and Trade); the Russian Science Foundation (grant 17-12-01029); the Spanish Ministerio de Economía y Competitividad (grants AYA2015-63939-C2-1-P, AYA2016-80889-P); the State Agency for Research of the Spanish MCIU through the "Center of Excellence Severo Ochoa" award for the Instituto de Astrofísica de Andalucía (SEV-2017-0709); the Toray Science Foundation; the US Department of Energy (USDOE) through the Los Alamos National Laboratory (operated by Triad National Security, LLC, for the National Nuclear Security Administration of the USDOE (Contract 89233218CNA000001)); the Italian Ministero dell'Istruzione Università e Ricerca through the grant Progetti Premiali 2012-iALMA (CUP C52I13000140001); the European Union's Horizon 2020 research and innovation programme under grant agreement No 730562 RadioNet; ALMA North America Development Fund; Chandra TM6-17006X.

This work used the Extreme Science and Engineering Discovery Environment (XSEDE), supported by NSF grant ACI-1548562, and CyVerse, supported by NSF grants DBI-0735191, DBI-1265383, and DBI-1743442. XSEDE Stampede2 resource at TACC was allocated through TG-AST170024 and TG-AST080026N. XSEDE JetStream resource at PTI and TACC was allocated through AST170028. The simulations were performed in part on the SuperMUC cluster at the LRZ in Garching, on the LOEWE cluster in CSC in Frankfurt, and on the HazelHen cluster at the HLRS in Stuttgart. This research was enabled in part by support provided by Compute Ontario (http://computeontario.ca), Calcul Quebec (http://www.calculquebec.ca) and Compute Canada (http://www.computecanada.ca).

We thank the staff at the participating observatories, correlation centers, and institutions for their enthusiastic support.

This Letter makes use of the following ALMA data: ADS/JAO.ALMA#2016.1.01154.V. ALMA is a partnership of the European Southern Observatory (ESO; Europe, representing its member states), NSF, and National Institutes of Natural Sciences of Japan, together with National Research Council (Canada), Ministry of Science and Technology (MOST; Taiwan), Academia Sinica Institute of Astronomy and Astrophysics (ASIAA; Taiwan), and Korea Astronomy and Space Science Institute (KASI; Republic of Korea), in cooperation with the Republic of Chile. The Joint ALMA Observatory is operated by ESO, Associated Universities, Inc. (AUI)/NRAO, and the National Astronomical Observatory of Japan (NAOJ). The NRAO is a facility of the NSF operated under cooperative agreement by AUI. APEX is a collaboration between the Max-Planck-Institut für Radioastronomie (Germany), ESO, and the Onsala Space Observatory (Sweden). The SMA is a joint project between the SAO and ASIAA and is funded by the Smithsonian Institution and the Academia Sinica. The JCMT is operated by the East Asian Observatory on behalf of the NAOJ, ASIAA, and KASI, as well as the Ministry of Finance of China, Chinese Academy of Sciences, and the National Key R&D Program (No. 2017YFA0402700) of China. Additional funding support for the JCMT is provided by the Science and Technologies Facility Council (UK) and participating universities in the UK and Canada. The LMT project is a joint effort of the Instituto Nacional de Astrófisica, Óptica, y Electrónica (Mexico) and the University of Massachusetts at Amherst (USA). The IRAM 30-m telescope on Pico Veleta, Spain is operated by IRAM and supported by CNRS (Centre National de la Recherche Scientifique, France), MPG (Max-Planck-Gesellschaft, Germany) and IGN (Instituto Geográfico Nacional, Spain).

The SMT is operated by the Arizona Radio Observatory, a part of the Steward Observatory of the University of Arizona, with financial support of operations from the State of Arizona and financial support for instrumentation development from the NSF. Partial SPT support is provided by the NSF Physics Frontier Center award (PHY-0114422) to the Kavli Institute of Cosmological Physics at the University of Chicago (USA), the Kavli Foundation, and the GBMF (GBMF-947). The SPT hydrogen maser was provided on loan from the GLT, courtesy of ASIAA. The SPT is supported by the National Science Foundation through grant PLR-1248097. Partial support is also provided by the NSF Physics Frontier Center grant PHY-1125897 to the Kavli Institute of Cosmological Physics at the University of Chicago, the Kavli Foundation and the Gordon and Betty Moore Foundation grant GBMF 947.

The EHTC has received generous donations of FPGA chips from Xilinx Inc., under the Xilinx University Program. The EHTC has benefited from technology shared under open-source license by the Collaboration for Astronomy Signal Processing and Electronics Research (CASPER). The EHT project is grateful to T4Science and Microsemi for their assistance with Hydrogen Masers. This research has made use of NASA's Astrophysics Data System. We gratefully acknowledge the support provided by the extended staff of the ALMA, both from the inception of the ALMA Phasing Project through the observational campaigns of 2017 and 2018. We would like to thank A. Deller and W. Brisken for EHT-specific support with the use of DiFX. We acknowledge the significance that Maunakea, where the SMA and JCMT EHT stations are located, has for the indigenous Hawaiian people.

## ORCID iDs

Kazunori Akiyama https://orcid.org/0000-0002-9475-4254
Antxon Alberdi https://orcid.org/0000-0002-9371-1033
Rebecca Azulay https://orcid.org/0000-0002-2200-5393
Anne-Kathrin Baczko https://orcid.org/0000-0003-3090-3975
Mislav Baloković https://orcid.org/0000-0003-0476-6647
John Barrett https://orcid.org/0000-0002-9290-0764
Lindy Blackburn https://orcid.org/0000-0002-9030-642X
Katherine L. Bouman https://orcid.org/0000-0003-0077-4367
Geoffrey C. Bower https://orcid.org/0000-0003-4056-9982
Christiaan D. Brinkerink https://orcid.org/0000-0002-2322-0749
Roger Brissenden https://orcid.org/0000-0002-2556-0894
Silke Britzen https://orcid.org/0000-0001-9240-6734
Avery E. Broderick https://orcid.org/0000-0002-3351-760X
Do-Young Byun https://orcid.org/0000-0003-1157-4109
Andrew Chael https://orcid.org/0000-0003-2966-6220
Chi-kwan Chan https://orcid.org/0000-0001-6337-6126
Shami Chatterjee https://orcid.org/0000-0002-2878-1502
Ilje Cho https://orcid.org/0000-0001-6083-7521
Pierre Christian https://orcid.org/0000-0001-6820-9941
John E. Conway https://orcid.org/0000-0003-2448-9181
Geoffrey B. Crew https://orcid.org/0000-0002-2079-3189
Yuzhu Cui https://orcid.org/0000-0001-6311-4345
Jordy Davelaar https://orcid.org/0000-0002-2685-2434
Mariafelicia De Laurentis https://orcid.org/0000-0002-9945-682X
Roger Deane https://orcid.org/0000-0003-1027-5043
Jessica Dempsey https://orcid.org/0000-0003-1269-9667
Gregory Desvignes https://orcid.org/0000-0003-3922-4055
Jason Dexter https://orcid.org/0000-0003-3903-0373
Sheperd S. Doeleman https://orcid.org/0000-0002-9031-0904
Ralph P. Eatough https://orcid.org/0000-0001-6196-4135
Heino Falcke https://orcid.org/0000-0002-2526-6724
Vincent L. Fish https://orcid.org/0000-0002-7128-9345
Raquel Fraga-Encinas https://orcid.org/0000-0002-5222-1361
José L. Gómez https://orcid.org/0000-0003-4190-7613
Peter Galison https://orcid.org/0000-0002-6429-3872
Charles F. Gammie https://orcid.org/0000-0001-7451-8935
Boris Georgiev https://orcid.org/0000-0002-3586-6424
Roman Gold https://orcid.org/0000-0003-2492-1966
Minfeng Gu (顾敏峰) https://orcid.org/0000-0002-4455-6946
Mark Gurwell https://orcid.org/0000-0003-0685-3621
Kazuhiro Hada https://orcid.org/0000-0001-6906-772X
Ronald Hesper https://orcid.org/0000-0003-1918-6098
Luis C. Ho (何子山) https://orcid.org/0000-0001-6947-5846
Mareki Honma https://orcid.org/0000-0003-4058-9000
Chih-Wei L. Huang https://orcid.org/0000-0001-5641-3953
Shiro Ikeda https://orcid.org/0000-0002-2462-1448
Sara Issaoun https://orcid.org/0000-0002-5297-921X
David J. James https://orcid.org/0000-0001-5160-4486
Michael Janssen https://orcid.org/0000-0001-8685-6544
Britton Jeter https://orcid.org/0000-0003-2847-1712
Wu Jiang (江悟) https://orcid.org/0000-0001-7369-3539
Michael D. Johnson https://orcid.org/0000-0002-4120-3029
Svetlana Jorstad https://orcid.org/0000-0001-6158-1708
Taehyun Jung https://orcid.org/0000-0001-7003-8643
Mansour Karami https://orcid.org/0000-0001-7387-9333
Ramesh Karuppusamy https://orcid.org/0000-0002-5307-2919
Tomohisa Kawashima https://orcid.org/0000-0001-8527-0496
Garrett K. Keating https://orcid.org/0000-0002-3490-146X
Mark Kettenis https://orcid.org/0000-0002-6156-5617
Jae-Young Kim https://orcid.org/0000-0001-8229-7183
Junhan Kim https://orcid.org/0000-0002-4274-9373
Motoki Kino https://orcid.org/0000-0002-2709-7338
Jun Yi Koay https://orcid.org/0000-0002-7029-6658
Patrick M. Koch https://orcid.org/0000-0003-2777-5861
Shoko Koyama https://orcid.org/0000-0002-3723-3372
Michael Kramer https://orcid.org/0000-0002-4175-2271
Carsten Kramer https://orcid.org/0000-0002-4908-4925
Thomas P. Krichbaum https://orcid.org/0000-0002-4892-9586
Tod R. Lauer https://orcid.org/0000-0003-3234-7247
Sang-Sung Lee https://orcid.org/0000-0002-6269-594X
Yan-Rong Li (李彦荣) https://orcid.org/0000-0001-5841-9179
Zhiyuan Li (李志远) https://orcid.org/0000-0003-0355-6437
Michael Lindqvist https://orcid.org/0000-0002-3669-0715
Kuo Liu https://orcid.org/0000-0002-2953-7376
Elisabetta Liuzzo https://orcid.org/0000-0003-0995-5201
Laurent Loinard https://orcid.org/0000-0002-5635-3345
Ru-Sen Lu (路如森) https://orcid.org/0000-0002-7692-7967
Nicholas R. MacDonald https://orcid.org/0000-0002-6684-8691
Jirong Mao (毛基荣) https://orcid.org/0000-0002-7077-7195
Sera Markoff https://orcid.org/0000-0001-9564-0876
Daniel P. Marrone https://orcid.org/0000-0002-2367-1080

Alan P. Marscher https://orcid.org/0000-0001-7396-3332
Iván Martí-Vidal https://orcid.org/0000-0003-3708-9611
Lynn D. Matthews https://orcid.org/0000-0002-3728-8082
Lia Medeiros https://orcid.org/0000-0003-2342-6728
Karl M. Menten https://orcid.org/0000-0001-6459-0669
Yosuke Mizuno https://orcid.org/0000-0002-8131-6730
Izumi Mizuno https://orcid.org/0000-0002-7210-6264
James M. Moran https://orcid.org/0000-0002-3882-4414
Kotaro Moriyama https://orcid.org/0000-0003-1364-3761
Monika Moscibrodzka https://orcid.org/0000-0002-4661-6332
Cornelia Müller https://orcid.org/0000-0002-2739-2994
Hiroshi Nagai https://orcid.org/0000-0003-0292-3645
Neil M. Nagar https://orcid.org/0000-0001-6920-662X
Masanori Nakamura https://orcid.org/0000-0001-6081-2420
Ramesh Narayan https://orcid.org/0000-0002-1919-2730
Iniyan Natarajan https://orcid.org/0000-0001-8242-4373
Chunchong Ni https://orcid.org/0000-0003-1361-5699
Aristeidis Noutsos https://orcid.org/0000-0002-4151-3860
Héctor Olivares https://orcid.org/0000-0001-6833-7580
Gisela N. Ortiz-León https://orcid.org/0000-0002-2863-676X
Daniel C. M. Palumbo https://orcid.org/0000-0002-7179-3816
Ue-Li Pen https://orcid.org/0000-0003-2155-9578
Dominic W. Pesce https://orcid.org/0000-0002-5278-9221
Oliver Porth https://orcid.org/0000-0002-4584-2557
Ben Prather https://orcid.org/0000-0002-0393-7734
Jorge A. Preciado-López https://orcid.org/0000-0002-4146-0113
Hung-Yi Pu https://orcid.org/0000-0001-9270-8812
Venkatessh Ramakrishnan https://orcid.org/0000-0002-9248-086X
Ramprasad Rao https://orcid.org/0000-0002-1407-7944
Alexander W. Raymond https://orcid.org/0000-0002-5779-4767
Luciano Rezzolla https://orcid.org/0000-0002-1330-7103
Bart Ripperda https://orcid.org/0000-0002-7301-3908
Freek Roelofs https://orcid.org/0000-0001-5461-3687
Eduardo Ros https://orcid.org/0000-0001-9503-4892
Mel Rose https://orcid.org/0000-0002-2016-8746
Alan L. Roy https://orcid.org/0000-0002-1931-0135
Chet Ruszczyk https://orcid.org/0000-0001-7278-9707
Benjamin R. Ryan https://orcid.org/0000-0001-8939-4461
Kazi L. J. Rygl https://orcid.org/0000-0003-4146-9043
David Sánchez-Arguelles https://orcid.org/0000-0002-7344-9920
Mahito Sasada https://orcid.org/0000-0001-5946-9960
Tuomas Savolainen https://orcid.org/0000-0001-6214-1085
Lijing Shao https://orcid.org/0000-0002-1334-8853
Zhiqiang Shen (沈志强) https://orcid.org/0000-0003-3540-8746
Des Small https://orcid.org/0000-0003-3723-5404
Bong Won Sohn https://orcid.org/0000-0002-4148-8378
Jason SooHoo https://orcid.org/0000-0003-1938-0720
Fumie Tazaki https://orcid.org/0000-0003-0236-0600
Paul Tiede https://orcid.org/0000-0003-3826-5648
Remo P. J. Tilanus https://orcid.org/0000-0002-6514-553X
Michael Titus https://orcid.org/0000-0002-3423-4505
Kenji Toma https://orcid.org/0000-0002-7114-6010
Pablo Torne https://orcid.org/0000-0001-8700-6058
Sascha Trippe https://orcid.org/0000-0003-0465-1559
Ilse van Bemmel https://orcid.org/0000-0001-5473-2950
Huib Jan van Langevelde https://orcid.org/0000-0002-0230-5946
Daniel R. van Rossum https://orcid.org/0000-0001-7772-6131
John Wardle https://orcid.org/0000-0002-8960-2942
Jonathan Weintroub https://orcid.org/0000-0002-4603-5204
Norbert Wex https://orcid.org/0000-0003-4058-2837
Robert Wharton https://orcid.org/0000-0002-7416-5209
Maciek Wielgus https://orcid.org/0000-0002-8635-4242
George N. Wong https://orcid.org/0000-0001-6952-2147
Qingwen Wu (吴庆文) https://orcid.org/0000-0003-4773-4987
Ken Young https://orcid.org/0000-0002-3666-4920
André Young https://orcid.org/0000-0003-0000-2682
Ziri Younsi https://orcid.org/0000-0001-9283-1191
Feng Yuan (袁峰) https://orcid.org/0000-0003-3564-6437
J. Anton Zensus https://orcid.org/0000-0001-7470-3321
Guangyao Zhao https://orcid.org/0000-0002-4417-1659
Shan-Shan Zhao https://orcid.org/0000-0002-9774-3606
Juan-Carlos Algaba https://orcid.org/0000-0001-6993-1696
Jadyn Anczarski https://orcid.org/0000-0003-4317-3385
Uwe Bach https://orcid.org/0000-0002-7722-8412
Frederick K. Baganoff https://orcid.org/0000-0003-3852-6545
Bradford A. Benson https://orcid.org/0000-0002-5108-6823
Jay M. Blanchard https://orcid.org/0000-0002-2756-395X
Iain M. Coulson https://orcid.org/0000-0002-7316-4626
Thomas M. Crawford https://orcid.org/0000-0001-9000-5013
Sergio A. Dzib https://orcid.org/0000-0001-6010-6200
Andreas Eckart https://orcid.org/0000-0001-6049-3132
Wendeline B. Everett https://orcid.org/0000-0002-5370-6651
Joseph R. Farah https://orcid.org/0000-0003-4914-5625
Christopher H. Greer https://orcid.org/0000-0002-9590-0508
Daryl Haggard https://orcid.org/0000-0001-6803-2138
Nils W. Halverson https://orcid.org/0000-0003-2606-9340
Antonio Hernández-Gómez https://orcid.org/0000-0001-7520-4305
Rubén Herrero-Illana https://orcid.org/0000-0002-7758-8717
Atish Kamble https://orcid.org/0000-0003-0861-5168
Ryan Keisler https://orcid.org/0000-0002-5922-1137
Yusuke Kono https://orcid.org/0000-0002-4187-8747
Erik M. Leitch https://orcid.org/0000-0001-8553-9336
Kyle D. Massingill https://orcid.org/0000-0002-0830-2033
Hugo Messias https://orcid.org/0000-0002-2985-7994
Daniel Michalik https://orcid.org/0000-0002-7618-6556
Andrew Nadolski https://orcid.org/0000-0001-9479-9957
Joseph Neilsen https://orcid.org/0000-0002-8247-786X
Chi H. Nguyen https://orcid.org/0000-0001-9368-3186
Michael A. Nowak https://orcid.org/0000-0001-6923-1315
Harriet Parsons https://orcid.org/0000-0002-6327-3423
Scott N. Paine https://orcid.org/0000-0003-4622-5857
Rurik A. Primiani https://orcid.org/0000-0003-3910-7529
Alexandra S. Rahlin https://orcid.org/0000-0003-3953-1776

Pim Schellart https://orcid.org/0000-0002-8324-0880
Hotaka Shiokawa https://orcid.org/0000-0002-8847-5275
David R. Smith https://orcid.org/0000-0003-0692-8582
Anthony A. Stark https://orcid.org/0000-0002-2718-9996
Sjoerd T. Timmer https://orcid.org/0000-0003-0223-9368
Nathan Whitehorn https://orcid.org/0000-0002-3157-0407
Jan G. A. Wouterloot https://orcid.org/0000-0002-4694-6905
Melvin Wright https://orcid.org/0000-0002-9154-2440
Paul Yamaguchi https://orcid.org/0000-0002-6017-8199
Shuo Zhang https://orcid.org/0000-0002-2967-790X
Lucy Ziurys https://orcid.org/0000-0002-1805-3886

---

The Event Horizon Telescope Collaboration,
Kazunori Akiyama[1,2,3,4], Antxon Alberdi[5], Walter Alef[6], Keiichi Asada[7], Rebecca Azulay[8,9,6], Anne-Kathrin Baczko[6], David Ball[10], Mislav Baloković[4,11], John Barrett[2], Dan Bintley[12], Lindy Blackburn[4,11], Wilfred Boland[13], Katherine L. Bouman[4,11,14], Geoffrey C. Bower[15], Michael Bremer[16], Christiaan D. Brinkerink[17], Roger Brissenden[4,11], Silke Britzen[6], Avery E. Broderick[18,19,20], Dominique Broguiere[16], Thomas Bronzwaer[17], Do-Young Byun[21,22], John E. Carlstrom[23,24,25,26], Andrew Chael[4,11], Chi-kwan Chan[10,27], Shami Chatterjee[28], Koushik Chatterjee[29], Ming-Tang Chen[15], Yongjun Chen (陈永军)[30,31], Ilje Cho[21,22], Pierre Christian[10,11], John E. Conway[32], James M. Cordes[28], Geoffrey B. Crew[2], Yuzhu Cui[33,34], Jordy Davelaar[17], Mariafelicia De Laurentis[35,36,37], Roger Deane[38,39], Jessica Dempsey[12], Gregory Desvignes[6], Jason Dexter[40], Sheperd S. Doeleman[4,11], Ralph P. Eatough[6], Heino Falcke[17], Vincent L. Fish[2], Ed Fomalont[1], Raquel Fraga-Encinas[17], William T. Freeman[41,42], Per Friberg[12], Christian M. Fromm[36], José L. Gómez[5], Peter Galison[4,43,44], Charles F. Gammie[45,46], Roberto García[16], Olivier Gentaz[16], Boris Georgiev[19,20], Ciriaco Goddi[17,47], Roman Gold[36], Minfeng Gu (顾敏峰)[30,48], Mark Gurwell[11], Kazuhiro Hada[33,34], Michael H. Hecht[2], Ronald Hesper[49], Luis C. Ho (何子山)[50,51], Paul Ho[7], Mareki Honma[33,34], Chih-Wei L. Huang[7], Lei Huang (黄磊)[30,48], David H. Hughes[52], Shiro Ikeda[3,53,54,55], Makoto Inoue[7], Sara Issaoun[17],

David J. James[4,11], Buell T. Jannuzi[10], Michael Janssen[17], Britton Jeter[19,20], Wu Jiang (江悟)[30], Michael D. Johnson[4,11], Svetlana Jorstad[56,57], Taehyun Jung[21,22], Mansour Karami[18,19], Ramesh Karuppusamy[6], Tomohisa Kawashima[3], Garrett K. Keating[11], Mark Kettenis[58], Jae-Young Kim[6], Junhan Kim[10], Jongsoo Kim[21], Motoki Kino[3,59], Jun Yi Koay[7], Patrick M. Koch[7], Shoko Koyama[7], Michael Kramer[6], Carsten Kramer[16], Thomas P. Krichbaum[6], Cheng-Yu Kuo[60], Tod R. Lauer[61], Sang-Sung Lee[21], Yan-Rong Li (李彦荣)[62], Zhiyuan Li (李志远)[63,64], Michael Lindqvist[32], Kuo Liu[6], Elisabetta Liuzzo[65], Wen-Ping Lo[7,66], Andrei P. Lobanov[6], Laurent Loinard[67,68], Colin Lonsdale[2], Ru-Sen Lu (路如森)[30,6], Nicholas R. MacDonald[6], Jirong Mao (毛基荣)[69,70,71], Sera Markoff[29,72], Daniel P. Marrone[10], Alan P. Marscher[56], Iván Martí-Vidal[32,73], Satoki Matsushita[7], Lynn D. Matthews[2], Lia Medeiros[10,74], Karl M. Menten[6], Yosuke Mizuno[36], Izumi Mizuno[12], James M. Moran[4,11], Kotaro Moriyama[33,2], Monika Moscibrodzka[17], Cornelia Müller[6,17], Hiroshi Nagai[3,34], Neil M. Nagar[75], Masanori Nakamura[7], Ramesh Narayan[4,11], Gopal Narayanan[76], Iniyan Natarajan[39], Roberto Neri[16], Chunchong Ni[19,20], Aristeidis Noutsos[6], Hiroki Okino[33,77], Héctor Olivares[36], Gisela N. Ortiz-León[6], Tomoaki Oyama[33], Feryal Özel[10], Daniel C. M. Palumbo[4,11], Nimesh Patel[11], Ue-Li Pen[18,78,79,80], Dominic W. Pesce[4,11], Vincent Piétu[16], Richard Plambeck[81], Aleksandar PopStefanija[76], Oliver Porth[29,36], Ben Prather[45], Jorge A. Preciado-López[18], Dimitrios Psaltis[10], Hung-Yi Pu[18], Venkatessh Ramakrishnan[75], Ramprasad Rao[15], Mark G. Rawlings[12], Alexander W. Raymond[4,11], Luciano Rezzolla[36], Bart Ripperda[36], Freek Roelofs[17], Alan Rogers[2], Eduardo Ros[6], Mel Rose[10], Arash Roshanineshat[10], Helge Rottmann[6], Alan L. Roy[6], Chet Ruszczyk[2], Benjamin R. Ryan[82,83], Kazi L. J. Rygl[65], Salvador Sánchez[84], David Sánchez-Arguelles[52,85], Mahito Sasada[33,86], Tuomas Savolainen[6,87,88], F. Peter Schloerb[76], Karl-Friedrich Schuster[16], Lijing Shao[6,51], Zhiqiang Shen (沈志强)[30,31], Des Small[58], Bong Won Sohn[21,22,89], Jason SooHoo[2], Fumie Tazaki[33], Paul Tiede[19,20], Remo P. J. Tilanus[17,47,90], Michael Titus[2], Kenji Toma[91,92], Pablo Torne[6,84], Tyler Trent[10], Sascha Trippe[93], Shuichiro Tsuda[33], Ilse van Bemmel[58], Huib Jan van Langevelde[58,94], Daniel R. van Rossum[17], Jan Wagner[6], John Wardle[95], Jonathan Weintroub[4,11], Norbert Wex[6], Robert Wharton[6], Maciek Wielgus[4,11], George N. Wong[45], Qingwen Wu (吴庆文)[96], Ken Young[11], André Young[17], Ziri Younsi[97,36], Feng Yuan (袁峰)[30,48,98], Ye-Fei Yuan (袁业飞)[99], J. Anton Zensus[6], Guangyao Zhao[21], Shan-Shan Zhao[17,63], Ziyan Zhu[44], Juan-Carlos Algaba[7,100], Alexander Allardi[101], Rodrigo Amestica[102], Jadyn Anczarski[103], Uwe Bach[6], Frederick K. Baganoff[104], Christopher Beaudoin[2], Bradford A. Benson[26,24], Ryan Berthold[12], Jay M. Blanchard[75,58], Ray Blundell[11], Sandra Bustamente[105], Roger Cappallo[2], Edgar Castillo-Domínguez[105,106], Chih-Cheng Chang[7,107], Shu-Hao Chang[7], Song-Chu Chang[107], Chung-Chen Chen[7], Ryan Chilson[15], Tim C. Chuter[12], Rodrigo Córdova Rosado[4,11], Iain M. Coulson[12], Thomas M. Crawford[24,25], Joseph Crowley[108], John David[84], Mark Derome[2], Matthew Dexter[109], Sven Dornbusch[6], Kevin A. Dudevoir[2,144], Sergio A. Dzib[6], Andreas Eckart[6,110], Chris Eckert[2], Neal R. Erickson[76], Wendeline B. Everett[111], Aaron Faber[112], Joseph R. Farah[4,11,113], Vernon Fath[76], Thomas W. Folkers[10], David C. Forbes[10], Robert Freund[10], Arturo I. Gómez-Ruiz[105,106], David M. Gale[105], Feng Gao[30,40], Gertie Geertsema[114], David A. Graham[6], Christopher H. Greer[10], Ronald Grosslein[76], Frédéric Gueth[16], Daryl Haggard[115,116,117], Nils W. Halverson[118], Chih-Chiang Han[7], Kuo-Chang Han[107], Jinchi Hao[107], Yutaka Hasegawa[7], Jason W. Henning[23,119], Antonio Hernández-Gómez[67,120], Rubén Herrero-Illana[121], Stefan Heyminck[6], Akihiko Hirota[3,7], James Hoge[12], Yau-De Huang[7], C. M. Violette Impellizzeri[7,1], Homin Jiang[7], Atish Kamble[4,11], Ryan Keisler[25], Kimihiro Kimura[7], Yusuke Kono[3], Derek Kubo[122], John Kuroda[12], Richard Lacasse[102], Robert A. Laing[123], Erik M. Leitch[23], Chao-Te Li[7], Lupin C.-C. Lin[7,124], Ching-Tang Liu[107], Kuan-Yu Liu[7], Li-Ming Lu[107], Ralph G. Marson[125], Pierre L. Martin-Cocher[7], Kyle D. Massingill[10], Callie Matulonis[12], Martin P. McColl[10], Stephen R. McWhirter[2], Hugo Messias[121,126], Zheng Meyer-Zhao[7,127], Daniel Michalik[128,129], Alfredo Montaña[105,106], William Montgomerie[12], Matias Mora-Klein[102], Dirk Muders[6], Andrew Nadolski[46], Santiago Navarro[84], Joseph Neilsen[103], Chi H. Nguyen[10,130], Hiroaki Nishioka[7], Timothy Norton[11], Michael A. Nowak[131], George Nystrom[15], Hideo Ogawa[132], Peter Oshiro[15], Tomoaki Oyama[133], Harriet Parsons[12], Scott N. Paine[11], Juan Peñalver[84], Neil M. Phillips[121,126], Michael Poirier[2], Nicolas Pradel[7], Rurik A. Primiani[134], Philippe A. Raffin[15], Alexandra S. Rahlin[23,135], George Reiland[10], Christopher Risacher[16], Ignacio Ruiz[84], Alejandro F. Sáez-Madaín[102,126], Remi Sassella[16], Pim Schellart[17,136], Paul Shaw[7], Kevin M. Silva[12], Hotaka Shiokawa[11], David R. Smith[137,138], William Snow[15], Kamal Souccar[76], Don Sousa[2], T. K. Sridharan[11], Ranjani Srinivasan[15], William Stahm[12], Anthony A. Stark[11], Kyle Story[139], Sjoerd T. Timmer[17], Laura Vertatschitsch[11,134], Craig Walther[12], Ta-Shun Wei[7], Nathan Whitehorn[140], Alan R. Whitney[2], David P. Woody[141], Jan G. A. Wouterloot[12], Melvin Wright[142], Paul Yamaguchi[11], Chen-Yu Yu[7], Milagros Zeballos[105,143], Shuo Zhang[104], and Lucy Ziurys[10]

[1] National Radio Astronomy Observatory, 520 Edgemont Road, Charlottesville, VA 22903, USA
[2] Massachusetts Institute of Technology, Haystack Observatory, 99 Millstone Road, Westford, MA 01886, USA
[3] National Astronomical Observatory of Japan, 2-21-1 Osawa, Mitaka, Tokyo 181-8588, Japan
[4] Black Hole Initiative at Harvard University, 20 Garden Street, Cambridge, MA 02138, USA
[5] Instituto de Astrofísica de Andalucía-CSIC, Glorieta de la Astronomía s/n, E-18008 Granada, Spain
[6] Max-Planck-Institut für Radioastronomie, Auf dem Hügel 69, D-53121 Bonn, Germany

[7] Institute of Astronomy and Astrophysics, Academia Sinica, 11F of Astronomy-Mathematics Building, AS/NTU No. 1, Sec. 4, Roosevelt Rd, Taipei 10617, Taiwan, R.O.C.
[8] Departament d'Astronomia i Astrofísica, Universitat de València, C. Dr. Moliner 50, E-46100 Burjassot, València, Spain
[9] Observatori Astronòmic, Universitat de València, C. Catedrático José Beltrán 2, E-46980 Paterna, València, Spain
[10] Steward Observatory and Department of Astronomy, University of Arizona, 933 N. Cherry Ave., Tucson, AZ 85721, USA
[11] Center for Astrophysics | Harvard & Smithsonian, 60 Garden Street, Cambridge, MA 02138, USA
[12] East Asian Observatory, 660 N. A'ohoku Pl., Hilo, HI 96720, USA
[13] Nederlandse Onderzoekschool voor Astronomie (NOVA), PO Box 9513, 2300 RA Leiden, The Netherlands
[14] California Institute of Technology, 1200 East California Boulevard, Pasadena, CA 91125, USA
[15] Institute of Astronomy and Astrophysics, Academia Sinica, 645 N. A'ohoku Place, Hilo, HI 96720, USA
[16] Institut de Radioastronomie Millimétrique, 300 rue de la Piscine, F-38406 Saint Martin d'Hères, France
[17] Department of Astrophysics, Institute for Mathematics, Astrophysics and Particle Physics (IMAPP), Radboud University, P.O. Box 9010, 6500 GL Nijmegen, The Netherlands
[18] Perimeter Institute for Theoretical Physics, 31 Caroline Street North, Waterloo, ON, N2L 2Y5, Canada
[19] Department of Physics and Astronomy, University of Waterloo, 200 University Avenue West, Waterloo, ON, N2L 3G1, Canada
[20] Waterloo Centre for Astrophysics, University of Waterloo, Waterloo, ON N2L 3G1 Canada
[21] Korea Astronomy and Space Science Institute, Daedeok-daero 776, Yuseong-gu, Daejeon 34055, Republic of Korea
[22] University of Science and Technology, Gajeong-ro 217, Yuseong-gu, Daejeon 34113, Republic of Korea
[23] Kavli Institute for Cosmological Physics, University of Chicago, 5640 South Ellis Avenue, Chicago, IL 60637, USA
[24] Department of Astronomy and Astrophysics, University of Chicago, 5640 South Ellis Avenue, Chicago, IL 60637, USA
[25] Department of Physics, University of Chicago, 5720 South Ellis Avenue, Chicago, IL 60637, USA
[26] Enrico Fermi Institute, University of Chicago, 5640 South Ellis Avenue, Chicago, IL 60637, USA
[27] Data Science Institute, University of Arizona, 1230 N. Cherry Ave., Tucson, AZ 85721, USA
[28] Cornell Center for Astrophysics and Planetary Science, Cornell University, Ithaca, NY 14853, USA
[29] Anton Pannekoek Institute for Astronomy, University of Amsterdam, Science Park 904, 1098 XH, Amsterdam, The Netherlands
[30] Shanghai Astronomical Observatory, Chinese Academy of Sciences, 80 Nandan Road, Shanghai 200030, People's Republic of China
[31] Key Laboratory of Radio Astronomy, Chinese Academy of Sciences, Nanjing 210008, People's Republic of China
[32] Department of Space, Earth and Environment, Chalmers University of Technology, Onsala Space Observatory, SE-439 92 Onsala, Sweden
[33] Mizusawa VLBI Observatory, National Astronomical Observatory of Japan, 2-12 Hoshigaoka, Mizusawa, Oshu, Iwate 023-0861, Japan
[34] Department of Astronomical Science, The Graduate University for Advanced Studies (SOKENDAI), 2-21-1 Osawa, Mitaka, Tokyo 181-8588, Japan
[35] Dipartimento di Fisica "E. Pancini," Universitá di Napoli "Federico II," Compl. Univ. di Monte S. Angelo, Edificio G, Via Cinthia, I-80126, Napoli, Italy
[36] Institut für Theoretische Physik, Goethe-Universität Frankfurt, Max-von-Laue-Straße 1, D-60438 Frankfurt am Main, Germany
[37] INFN Sez. di Napoli, Compl. Univ. di Monte S. Angelo, Edificio G, Via Cinthia, I-80126, Napoli, Italy
[38] Department of Physics, University of Pretoria, Lynnwood Road, Hatfield, Pretoria 0083, South Africa
[39] Centre for Radio Astronomy Techniques and Technologies, Department of Physics and Electronics, Rhodes University, Grahamstown 6140, South Africa
[40] Max-Planck-Institut für Extraterrestrische Physik, Giessenbachstr. 1, D-85748 Garching, Germany
[41] Department of Electrical Engineering and Computer Science, Massachusetts Institute of Technology, 32-D476, 77 Massachussetts Ave., Cambridge, MA 02142, USA
[42] Google Research, 355 Main St., Cambridge, MA 02142, USA
[43] Department of History of Science, Harvard University, Cambridge, MA 02138, USA
[44] Department of Physics, Harvard University, Cambridge, MA 02138, USA
[45] Department of Physics, University of Illinois, 1110 West Green St, Urbana, IL 61801, USA
[46] Department of Astronomy, University of Illinois at Urbana-Champaign, 1002 West Green Street, Urbana, IL 61801, USA
[47] Leiden Observatory—Allegro, Leiden University, P.O. Box 9513, 2300 RA Leiden, The Netherlands
[48] Key Laboratory for Research in Galaxies and Cosmology, Chinese Academy of Sciences, Shanghai 200030, People's Republic of China
[49] NOVA Sub-mm Instrumentation Group, Kapteyn Astronomical Institute, University of Groningen, Landleven 12, 9747 AD Groningen, The Netherlands
[50] Department of Astronomy, School of Physics, Peking University, Beijing 100871, People's Republic of China
[51] Kavli Institute for Astronomy and Astrophysics, Peking University, Beijing 100871, People's Republic of China
[52] Instituto Nacional de Astrofísica, Óptica y Electrónica. Apartado Postal 51 y 216, 72000. Puebla Pue., México
[53] The Institute of Statistical Mathematics, 10-3 Midori-cho, Tachikawa, Tokyo, 190-8562, Japan
[54] Department of Statistical Science, The Graduate University for Advanced Studies (SOKENDAI), 10-3 Midori-cho, Tachikawa, Tokyo 190-8562, Japan
[55] Kavli Institute for the Physics and Mathematics of the Universe, The University of Tokyo, 5-1-5 Kashiwanoha, Kashiwa, 277-8583, Japan
[56] Institute for Astrophysical Research, Boston University, 725 Commonwealth Ave., Boston, MA 02215, USA
[57] Astronomical Institute, St. Petersburg University, Universitetskij pr., 28, Petrodvorets, 198504 St. Petersburg, Russia
[58] Joint Institute for VLBI ERIC (JIVE), Oude Hoogeveensedijk 4, 7991 PD Dwingeloo, The Netherlands
[59] Kogakuin University of Technology & Engineering, Academic Support Center, 2665-1 Nakano, Hachioji, Tokyo 192-0015, Japan
[60] Physics Department, National Sun Yat-Sen University, No. 70, Lien-Hai Rd, Kaosiung City 80424, Taiwan, R.O.C
[61] National Optical Astronomy Observatory, 950 North Cherry Ave., Tucson, AZ 85719, USA
[62] Key Laboratory for Particle Astrophysics, Institute of High Energy Physics, Chinese Academy of Sciences, 19B Yuquan Road, Shijingshan District, Beijing, People's Republic of China
[63] School of Astronomy and Space Science, Nanjing University, Nanjing 210023, People's Republic of China
[64] Key Laboratory of Modern Astronomy and Astrophysics, Nanjing University, Nanjing 210023, People's Republic of China
[65] Italian ALMA Regional Centre, INAF-Istituto di Radioastronomia, Via P. Gobetti 101, I-40129 Bologna, Italy
[66] Department of Physics, National Taiwan University, No.1, Sect.4, Roosevelt Rd., Taipei 10617, Taiwan, R.O.C
[67] Instituto de Radioastronomía y Astrofísica, Universidad Nacional Autónoma de México, Morelia 58089, México
[68] Instituto de Astronomía, Universidad Nacional Autónoma de México, CdMx 04510, México
[69] Yunnan Observatories, Chinese Academy of Sciences, 650011 Kunming, Yunnan Province, People's Republic of China
[70] Center for Astronomical Mega-Science, Chinese Academy of Sciences, 20A Datun Road, Chaoyang District, Beijing, 100012, People's Republic of China
[71] Key Laboratory for the Structure and Evolution of Celestial Objects, Chinese Academy of Sciences, 650011 Kunming, People's Republic of China
[72] Gravitation Astroparticle Physics Amsterdam (GRAPPA) Institute, University of Amsterdam, Science Park 904, 1098 XH Amsterdam, The Netherlands
[73] Centro Astronómico de Yebes (IGN), Apartado 148, E-19180 E-Yebes, Spain
[74] Department of Physics, Broida Hall, University of California Santa Barbara, Santa Barbara, CA 93106, USA
[75] Astronomy Department, Universidad de Concepción, Casilla 160-C, Concepción, Chile
[76] Department of Astronomy, University of Massachusetts, 01003, Amherst, MA, USA
[77] Department of Astronomy, Graduate School of Science, The University of Tokyo, 7-3-1 Hongo, Bunkyo-ku, Tokyo 113-0033, Japan

[78] Canadian Institute for Theoretical Astrophysics, University of Toronto, 60 St. George Street, Toronto, ON M5S 3H8, Canada
[79] Dunlap Institute for Astronomy and Astrophysics, University of Toronto, 50 St. George Street, Toronto, ON M5S 3H4, Canada
[80] Canadian Institute for Advanced Research, 180 Dundas St West, Toronto, ON M5G 1Z8, Canada
[81] Radio Astronomy Laboratory, University of California, Berkeley, CA 94720, USA
[82] CCS-2, Los Alamos National Laboratory, P.O. Box 1663, Los Alamos, NM 87545, USA
[83] Center for Theoretical Astrophysics, Los Alamos National Laboratory, Los Alamos, NM, 87545, USA
[84] Instituto de Radioastronomía Milimétrica, IRAM, Avenida Divina Pastora 7, Local 20, E-18012, Granada, Spain
[85] Consejo Nacional de Ciencia y Tecnología, Av. Insurgentes Sur 1582, 03940, Ciudad de México, México
[86] Hiroshima Astrophysical Science Center, Hiroshima University, 1-3-1 Kagamiyama, Higashi-Hiroshima, Hiroshima 739-8526, Japan
[87] Aalto University Department of Electronics and Nanoengineering, PL 15500, FI-00076 Aalto, Finland
[88] Aalto University Metsähovi Radio Observatory, Metsähovintie 114, FI-02540 Kylmälä, Finland
[89] Department of Astronomy, Yonsei University, Yonsei-ro 50, Seodaemun-gu, 03722 Seoul, Republic of Korea
[90] Netherlands Organisation for Scientific Research (NWO), Postbus 93138, 2509 AC Den Haag , The Netherlands
[91] Frontier Research Institute for Interdisciplinary Sciences, Tohoku University, Sendai 980-8578, Japan
[92] Astronomical Institute, Tohoku University, Sendai 980-8578, Japan
[93] Department of Physics and Astronomy, Seoul National University, Gwanak-gu, Seoul 08826, Republic of Korea
[94] Leiden Observatory, Leiden University, Postbus 2300, 9513 RA Leiden, The Netherlands
[95] Physics Department, Brandeis University, 415 South Street, Waltham, MA 02453, USA
[96] School of Physics, Huazhong University of Science and Technology, Wuhan, Hubei, 430074, People's Republic of China
[97] Mullard Space Science Laboratory, University College London, Holmbury St. Mary, Dorking, Surrey, RH5 6NT, UK
[98] School of Astronomy and Space Sciences, University of Chinese Academy of Sciences, No. 19A Yuquan Road, Beijing 100049, People's Republic of China
[99] Astronomy Department, University of Science and Technology of China, Hefei 230026, People's Republic of China
[100] Department of Physics, Faculty of Science, University of Malaya, 50603 Kuala Lumpur, Malaysia
[101] University of Vermont, Burlington, VT 05405, USA
[102] National Radio Astronomy Observatory, NRAO Technology Center, 1180 Boxwood Estate Road, Charlottesville, VA 22903, USA
[103] Department of Physics, Villanova University, 800 E. Lancaster Ave, Villanova, PA, 19085, USA
[104] Kavli Institute for Astrophysics and Space Research, Massachusetts Institute of Technology, Cambridge, MA 02139, USA
[105] Instituto Nacional de Astrofísica, Óptica y Electrónica, Luis Enrique Erro 1, Tonantzintla, Puebla, C.P. 72840, Mexico
[106] Consejo Nacional de Ciencia y Tecnología, Av. Insurgentes Sur 1582, Col. Crédito Constructor, CDMX, C.P. 03940, Mexico
[107] National Chung-Shan Institute of Science and Technology, No.566, Ln. 134, Longyuan Rd., Longtan Dist., Taoyuan City 325, Taiwan, R.O.C.
[108] MIT Haystack Observatory, 99 Millstone Road, Westford, MA 01886, USA
[109] Dept. of Astronomy, Univ. of California Berkeley, 501 Campbell, Berkeley, CA 94720, USA
[110] Physikalisches Institut der Universität zu Köln, Zülpicher Str. 77, D-50937 Köln, Germany
[111] CASA, Department of Astrophysical and Planetary Sciences, University of Colorado, Boulder, CO 80309, USA
[112] Western University, 1151 Richmond Street, London, Ontario, N6A 3K7, Canada
[113] University of Massachusetts Boston, 100 William T, Morrissey Blvd, Boston, MA 02125, USA
[114] Research and Development Weather and Climate Models, Royal Netherlands Meteorological Institute, Utrechtseweg 297, 3731 GA, De Bilt, The Netherlands
[115] Department of Physics, McGill University, 3600 University Street, Montréal, QC H3A 2T8, Canada
[116] McGill Space Institute, McGill University, 3550 University Street, Montréal, QC H3A 2A7, Canada
[117] CIFAR Azrieli Global Scholar, Gravity & the Extreme Universe Program, Canadian Institute for Advanced Research, 661 University Avenue, Suite 505, Toronto, ON M5G 1M1, Canada
[118] Department of Astrophysical and Planetary Sciences and Department of Physics, University of Colorado, Boulder, CO 80309, USA
[119] High Energy Physics Division, Argonne National Laboratory, 9700 S. Cass Avenue, Argonne, IL, 60439, USA
[120] IRAP, Université de Toulouse, CNRS, UPS, CNES, Toulouse, France
[121] European Southern Observatory, Alonso de Córdova 3107, Vitacura, Casilla 19001, Santiago de Chile, Chile
[122] ASIAA Hilo Office, 645 N. A'ohoku Place, University Park, Hilo, HI 96720, USA
[123] Square Kilometre Array Organisation, Jodrell Bank Observatory, Lower Withington, Macclesfield, Cheshire SK11 9DL, UK
[124] Department of Physics, Ulsan National Institute of Science and Technology, Ulsan, 44919, Republic of Korea
[125] National Radio Astronomy Observatory, P.O. Box O, Socorro, NM 87801, USA
[126] Joint ALMA Observatory, Alonso de Córdova 3107, Vitacura 763-0355, Santiago de Chile, Chile
[127] ASTRON, Oude Hoogeveensedijk 4, 7991 PD Dwingeloo, The Netherlands
[128] Science Support Office, Directorate of Science, European Space Research and Technology Centre (ESA/ESTEC), Keplerlaan 1, 2201 AZ Noordwijk, The Netherlands
[129] University of Chicago, 5640 South Ellis Avenue, Chicago, IL 60637, USA
[130] Center for Detectors, School of Physics and Astronomy, Rochester Institute of Technology, 1 Lomb Memorial Drive, Rochester, NY 14623, USA
[131] Physics Dept., CB 1105, Washington University, One Brookings Drive, St. Louis, MO 63130-4899, USA
[132] Osaka Prefecture University, Gakuencyou Sakai Osaka, Sakai 599-8531, Kinki, Japan
[133] Mizusawa VLBI Observatory, National Astronomical Observatory of Japan, Ohshu, Iwate 023-0861, Japan
[134] Systems & Technology Research, 600 West Cummings Park, Woburn, MA 01801, USA
[135] Fermi National Accelerator Laboratory, PO Box 500, Batavia, IL 60510, USA
[136] Department of Astrophysical Sciences, Princeton University, Princeton, NJ 08544, USA
[137] MERLAB, 357 S. Candler St., Decatur, GA 30030, USA
[138] GWW School of Mechanical Engineering, Georgia Institute of Technology, 771 Ferst Dr., Atlanta, GA 30332, USA
[139] Kavli Institute for Particle Astrophysics and Cosmology, Stanford University, 452 Lomita Mall, Stanford, CA 94305, USA
[140] Dept. of Physics and Astronomy, UCLA, Los Angeles, CA 90095, USA
[141] Owens Valley Radio Observatory, California Institute of Technology, Big Pine, CA 93513, USA
[142] Dept. of Astronomy, Radio Astronomy Laboratory, Univ. of California Berkeley, 601 Campbell, Berkeley, CA 94720, USA
[143] Universidad de las Américas Puebla, Sta. Catarina Mártir S/N, San Andrés Cholula, Puebla, C.P. 72810, Mexico
[144] Deceased.